\pdfoutput=1

\documentclass[aps,prd,twocolumn,preprintnumbers,superscriptaddress,floatfix,nofootinbib]{revtex4}

\usepackage{multirow,graphicx,amssymb,url,mathrsfs,amsmath,epstopdf}
\usepackage{comment,color}
\usepackage{hyperref}
\usepackage{rotating}
\usepackage[mathscr]{eucal}
\usepackage{eucal,wrapfig,boxedminipage,setspace,subfigure}
\usepackage{amsxtra,amstext,latexsym,dsfont}

               \def\D  {\Delta}

\def\IR{{\hbox{{\rm I}\kern-.2em\hbox{\rm R}}}}
\def\IB{{\hbox{{\rm I}\kern-.2em\hbox{\rm B}}}}
\def\IN{{\hbox{{\rm I}\kern-.2em\hbox{\rm N}}}}
\def\IC{\,\,{\hbox{{\rm I}\kern-.59em\hbox{\bf C}}}}
\def\IZ{{\hbox{{\rm Z}\kern-.4em\hbox{\rm Z}}}}
\def\IP{{\hbox{{\rm I}\kern-.2em\hbox{\rm P}}}}
\def\IH{{\hbox{{\rm I}\kern-.4em\hbox{\rm H}}}}
\def\ID{{\hbox{{\rm I}\kern-.2em\hbox{\rm D}}}}

\def\be{\begin{equation}}
\def\ee{\end{equation}}
\def\ba{\begin{eqnarray}}
\def\ea{\end{eqnarray}}

\def\half{\frac{1}{2}}

\def\nn{\nonumber}
\def\ea{{\it et al}. }

\newcommand{\beq}{\begin{equation}}
\newcommand{\eeq}{\end{equation}}
\newcommand{\bea}{\begin{eqnarray}}
\newcommand{\eea}{\end{eqnarray}}

\begin{document}

\newcommand\sect[1]{\emph{#1}---}

\pagestyle{empty}

\preprint{
\begin{minipage}[t]{3in}
\begin{flushright} SHEP-10-25
\\[30pt]
\hphantom{.}
\end{flushright}
\end{minipage}
}

\title{Non Mean-Field Quantum Critical Points from Holography}

\author{Nick Evans}
\email{evans@soton.ac.uk}
\affiliation{ School of Physics and Astronomy, University of
Southampton, Southampton, SO17 1BJ, UK \\ 
}
\author{Kristan Jensen}
\email{kristanj@u.washington.edu}
\affiliation{Department of Physics, University of Washington, 
Seattle, WA 98195-1560, USA 
 \\ 
 }
\author{Keun-Young Kim}
\email{k.kim@soton.ac.uk}
\affiliation{ School of Physics and Astronomy, University of
Southampton, Southampton, SO17 1BJ, UK \\ 
}


\begin{abstract}
We construct a class of quantum critical points with non-mean-field critical exponents via
 holography. Our approach is phenomenological. Beginning with the D3/D5 system at nonzero
 density and magnetic field which has a chiral phase transition, 
 we simulate the addition of a third control parameter. We
 then identify a line of quantum critical points in the phase diagram of this theory,
 provided that the simulated control parameter has dimension less than two. This line
 smoothly interpolates between a second-order transition with mean-field exponents at
 zero magnetic field to a holographic Berezinskii-Kosterlitz-Thouless transition at larger
 magnetic fields. The critical exponents of these transitions only depend upon the parameters
 of an emergent infrared theory. Moreover, the non-mean-field scaling is destroyed at any
 nonzero temperature. We discuss how generic these transitions are.

\end{abstract}
\maketitle

\section{Introduction  and summary }

The AdS/CFT  Correspondence
\cite{Maldacena:1997re,Witten:1998qj,Gubser:1998bc} describes a
class of strongly coupled gauge theories in terms of weakly
coupled gravitational systems. It has proved an extremely
versatile tool for the study of strong coupling phenomena over the
last ten years. For example, the correspondence has been used to
study the physics of deconfined plasmas, including
transport~\cite{Son:2007vk} and energy
loss~\cite{Herzog:2006gh,Gubser:2006bz}. Recently there has been
much interest in the use of AdS/CFT to realize condensed matter
phenomena. Much of this work has been dedicated to the study of
non-Fermi liquids~\cite{Lee:2008xf,Cubrovic:2009ye,Faulkner:2009wj} and
holographic superfluids~\cite{Hartnoll:2008vx,Gauntlett:2009dn} in
the hope of better understanding the phase diagram of
high-temperature superconductors. Another route to the same goal
involves the study of quantum critical points in strongly
interacting theories. These zero temperature transitions are
interesting in their own right, as they tend to govern the physics
of large swaths of the phase diagram at nonzero temperature.
Indeed, the ``strange metal'' phases observed in high-temperature
superconductors may originate from a quantum critical
point~\cite{sachdev-2009}.

The study of  critical phenomena is of central importance in the
condensed matter community. At any continuous phase transition
there is an emergent infrared fixed point~\cite{Wilson:1973jj}. Of
particular interest are transitions where the infrared theory is
itself an interacting quantum field theory. These transitions are
characterized by non-mean-field critical exponents. It would be
extremely interesting if holography can be used to study these
transitions or perhaps even transitions beyond the
Landau-Ginzburg-Wilson paradigm altogether~\cite{SVBSF}.
Unfortunately, most continuous transitions in holographic models
are second-order with mean-field exponents~\cite{Karch:2007br}. In
fact, the mean-field exponents should be expected rather than
surprising. They appear because of the large $N$ parameter in
these theories, which allows us to study them via their
holographic duals. In this limit quantum fluctuations are
suppressed in both the field~\cite{Yaffe:1981vf} and gravitational
theories. As a result, examples of non-mean-field exponents in
large $N$ theories are doubly interesting.  Moreover some
\emph{justification} for their non-mean-field behavior should be
given in the sense of~\cite{Rosenstein:1988dj}.

These questions are  not only useful for the condensed matter
community. The study of phase transitions in large $N$ theories
necessarily sheds light on the physics of non-Abelian gauge
theories. A general classification of transitions in the phase
diagram of such theories is important. Such a dictionary will help
in our understanding of QCD-like gauge theories in $(3+1)$
dimensions as well as in condensed matter systems in $(2+1)$
dimensions. Much is already known. For example, the finite
temperature deconfinement transitions of these gauge theories with
holographic duals are first order~\cite{Witten:1998zw} and map
onto Hawking-Page~\cite{Hawking:1982dh} phase transitions in the
gravity description. Additional transitions have been identified
in systems with
quarks~\cite{Karch:2002sh,Grana:2001xn,Bertolini:2001qa,Kruczenski:2003be,Erdmenger:2007cm,Karch:2000gx,DeWolfe:2001pq,Erdmenger:2002ex}
- there are meson-melting transitions in a
thermal~\cite{Babington:2003vm,Apreda:2005yz,Albash:2006ew,Mateos:2006nu,Mateos:2007vn,Hoyos:2006gb,Peeters:2006iu,Erdmenger:2007ja,Kaminski:2009ce}
or high-density bath~\cite{Nakamura:2006xk,Kobayashi:2006sb,Nakamura:2007nx,Mateos:2007vc}. Chiral
symmetry is also broken in the system with a magnetic
field~\cite{Filev:2007gb,Albash:2007bk,Filev:2007qu,Erdmenger:2007bn,Zayakin:2008cy,Filev:2009xp,Filev:2009ai,D'Hoker:2010rz,D'Hoker:2010ij}
and there is a chiral restoration transition at large densities
that occurs in addition to the meson-melting transition. All of
these transitions are typically first order at finite temperature
and low quark density but continuous at large density
(see~\cite{Evans:2010iy,Jensen:2010vd,Evans:2010hi} for the full
phase diagrams of the D3/D5 and D3/D7 systems with magnetic field
which we will study here. For thermodynamics see ~\cite{Wapler:2009rf,Wapler:2010nq}).

Flavored gauge theories  are also natural to study from the
critical phenomena perspective. Flavor sectors carry new
symmetries, leading to a richer phase diagram. At large $N$, we
also get the expansion parameter $N_f/N$ and so the quenched limit
is simultaneously rich and tractable. The holographic description
of these theories involves probe D-branes minimizing their
worldvolume in the dual geometry~\cite{Karch:2002sh}.

The first example of a holographic  quantum critical point
separating two nonzero density phases was obtained in the D3/D7
system~\cite{Evans:2010iy,Jensen:2010vd}. The dual field theory is
simply strongly coupled $\mathcal{N}=4$ super Yang-Mills (SYM)
coupled to a small number of massless fundamental hypermultiplets.
The chiral transition in this theory is triggered by large
magnetic fields and is second-order with mean-field critical
exponents. In fact, there are related chiral transitions at
nonzero density and magnetic field for almost all of the
supersymmetric probe brane systems~\cite{Jensen:2010vx}. Most of
these are second-order transitions with mean-field exponents, but
a handful are not.

In particular, the first example of a non-second-order,
non-mean-field transition in holography was identified in the
D3/D5 system~\cite{Jensen:2010ga}. The dual field theory is the
same as in the D3/D7 setup, but the flavor fields are confined to
a $(2+1)$ dimensional defect. The chiral transition in this theory
exhibits exponential scaling and so is reminiscent of the
celebrated Berezinskii-Kosterlitz-Thouless (BKT)
transition~\cite{Kosterlitz:1973xp,Kosterlitz:1974sm,Berezinskii}.
This is the first known instance of exponential scaling at zero
temperature in $(2+1)$ dimensions. Indeed, one of us helped term
this new transition a \emph{holographic} BKT transition, as it
occurs in a different context than the original BKT transition.
Since then holographic BKT transitions have also been found in the
context of extremal asymptotically AdS$_4$ dyonic black
holes~\cite{Iqbal:2010eh} as well as in two other probe brane
setups, namely flavored ABJM theory and flavored $(1,1)$ little
string theory~\cite{Jensen:2010vx,Pal:2010gj}. They have
also been identified in noncritical string setups in~\cite{Gursoy}.

The existence and properties of the BKT transition are  intimately
related to the Coleman-Mermin-Wagner
theorem~\cite{Coleman:1973ci,Mermin:1966fe,Hohenberg:1967zz}.
Recall that transitions of the BKT type are between disordered and
\emph{quasi}-ordered phases in two dimensions. In the
quasi-ordered phase, two-point functions of symmetry-breaking
operators have polynomial falloff at long distances while the
correlation length in the disordered phases scales as $\exp
(c/\sqrt{T-T_c})$ near the critical temperature
$T_c$~\cite{Kosterlitz:1974sm}. Holographic BKT transitions are
novel in that they exhibit exponential scaling in an
\emph{ordered} phase. Correspondingly,  their existence is
entirely unrelated to the long-distance restoration of continuous
symmetry in two dimensions. On the gravity side, holographic BKT
transitions occur due to the violation of the
Breitenlohner-Freedman (BF)~\cite{Breitenlohner:1982jf} bound in
the infrared region of the dual geometry by a scalar field dual to
an order parameter. In the field theory, this amounts to
considering a theory with an emergent CFT in the infrared. The
transition is driven by taking an operator dimension in the
emergent theory into the complex plane. On general grounds
presented in~\cite{Kaplan:2009kr} this violation was expected to
produce BKT scaling, but the D3/D5 system was the first example of
a setting where the BF bound is violated controllably.

In all known examples of holographic BKT transitions the dual
geometry has an effective infrared AdS$_2$ region. In the case of
extremal asymptotically AdS$_4$ dyonic black holes, the
near-horizon geometry is of the form AdS$_2\times
\mathbb{R}^2$~\cite{Faulkner:2009wj}. For the probe brane systems,
there is no physical AdS$_2$ region at the bottom of the geometry,
but worldvolume fields obey the equations of motion for fields in
an AdS$_2$-like region there~\cite{Seiberg:1999vs}. For the dyonic
black holes the emergent CFT is important: it governs much of the
low-frequency form of correlation functions at zero and small
temperatures~\cite{Faulkner:2009wj}. In this work, the emergent
CFT will also be of critical importance.

With all of this background in mind, there is an important open
question:  can we use the holographic technique to study
second-order quantum critical points with non-mean-field
exponents?\footnote{At nonzero temperature, there are two known
classes for these transitions. In the first, there is a curvature
singularity at the transition~\cite{Karch:2009ph,Buchel:2009mf}.
The second class is phenomenological, where the non-mean-field
exponents arise from an effective bulk action~\cite{Franco:2009if}
whose terms presumably originate from $1/N$ corrections.} If so,
can we explain the exponents at large $N$? In this work we answer
both questions in the affirmative\footnote{As we were finishing this work, the authors of~\cite{Faulkner:2010gj} released a paper that studies holographic superfluids in the presence of double-trace deformations. They obtain quantum critical points with non mean-field scalings that precisely match our results.}.

\begin{figure}[]
\includegraphics[width=8cm]{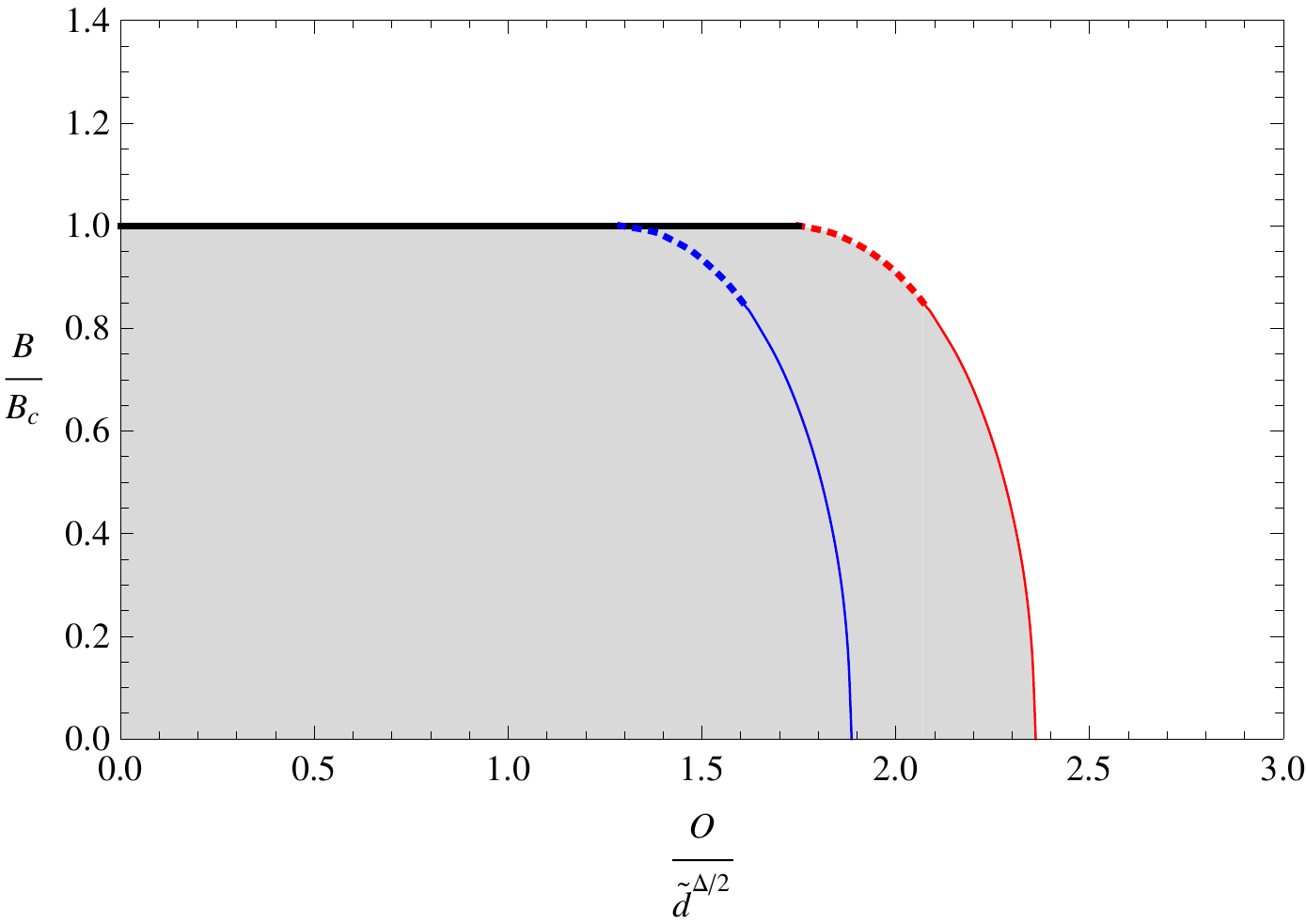}
\caption{\label{phases}The zero temperature phase diagram of our
phenomenologically deformed D3/D5 system at nonzero (fixed)
density, magnetic field, and simulated control parameter $O$ of
dimension $\Delta$. The shaded region indicates the chirally
symmetric phase and the white region the broken phase. For
$\Delta<2$, there is a line of holographic BKT transitions at the
critical magnetic field $B_c=\tilde{d}/\sqrt{7}$. There is also a
line of second-order quantum critical points triggered by $O$ that
connects to the line of BKT transitions. The position of the line
depends on the precise manner we introduce $O$. Indeed, the left
line corresponds to the choice $\Delta=5/4$ and the right to
$\Delta=1$. However, the critical exponents are only a function of
the magnetic field and density (Fig.\ref{nmfExp}). The static critical exponent
$\beta$ takes the mean-field value $1/2$ for the solid portion of
the lines, but takes on a non mean-field value for the dotted
section as in Eq.~(\ref{betaExp}). The line of critical points
then interpolates between a second-order mean-field transition and
a holographic BKT transition.}
\end{figure}

Our philosophy is to obtain the most general quantum critical
point  in a probe brane system (in the strict $N\rightarrow\infty$
limit). We do this by considering a theory with three relevant
control parameters at zero temperature, one that tends to preserve
a symmetry (nonzero density) and the other two to break it 
(one of them is a magnetic field). 
Ideally to explore this one should use explicit examples of such a theory. 
However, the gravitational description of the probe branes is rather
restrictive since only a very small number of operators have their
dimensions protected and appear as modes in the Dirac-Born-Infeld
(DBI) action of the brane - the rest are stringy modes and have
very large dimension. We therefore take a phenomenological
approach in this paper and begin with the D3/D5 system at nonzero
density and magnetic field. We simply include our third control
parameter, $O$, into the brane action by hand in a natural
fashion. It becomes a magnetic field when its dimension is two and
otherwise we tune its dimension. As long as $O$ has dimension less
than two\footnote{When $O$ has a larger dimension than the
density, chiral symmetry is broken for any nonzero $O$ as
in~\cite{Filev:2007gb}}, the resulting phase diagram of the theory
is quite rich; we plot it in Fig.~\ref{phases}. 
At fixed density
and magnetic field, we find a chiral quantum critical point as we
vary $O$. Varying the magnetic field  leads to a line of
second-order transitions that connects to a line of holographic
BKT transitions at critical magnetic fields.

\begin{figure}[]
\includegraphics[width=8cm]{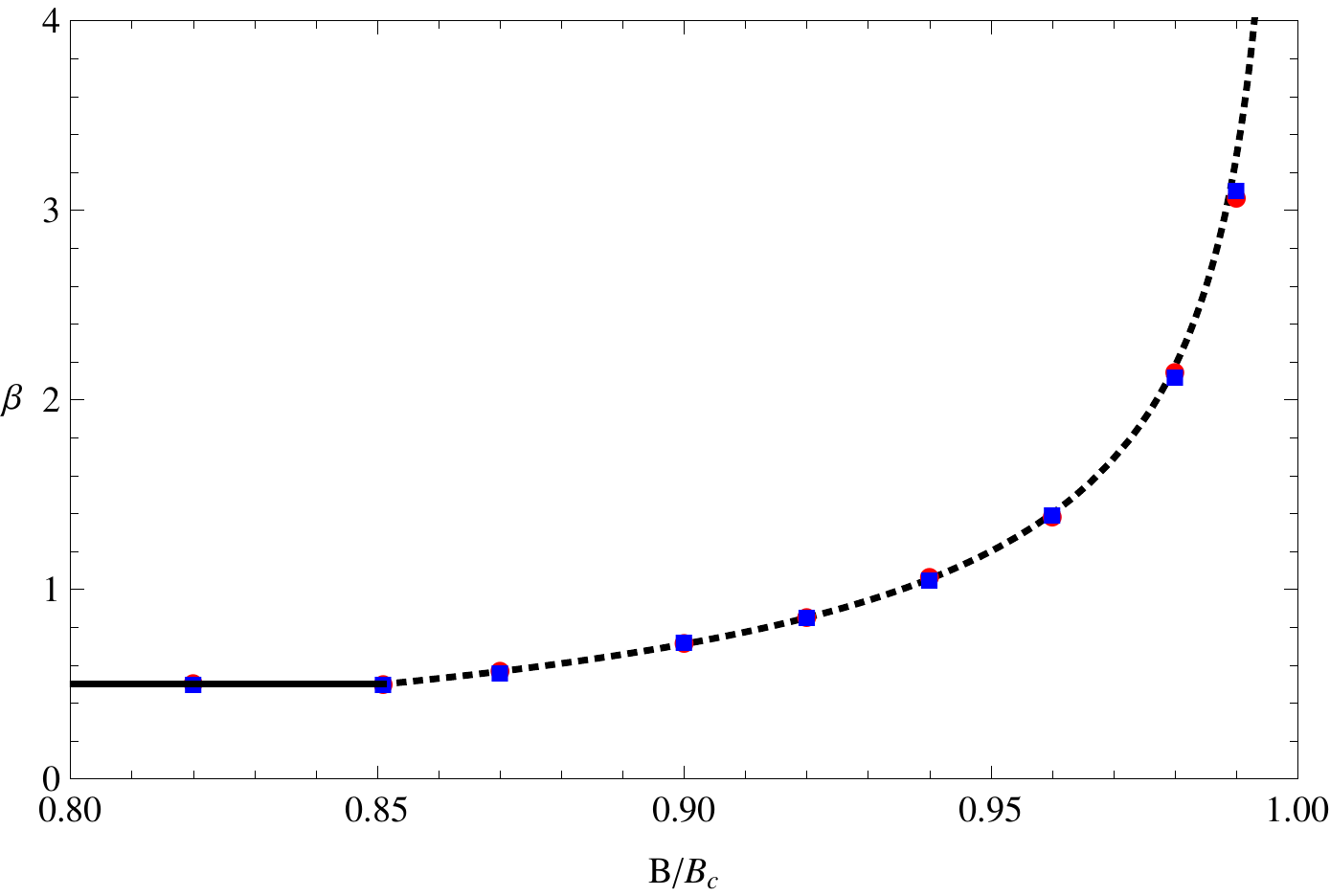}
\caption{\label{nmfExp}The critical exponent $\beta$ in the
deformed D3/D5 system as a function of magnetic field at
$\Delta=1$ and $\Delta=5/4$. For small magnetic fields - such that
$\Delta_{\rm IR}>3/4$ - the exponent $\beta$ assumes a mean-field
value, while for larger magnetic fields it does not. 
The exponents for $\Delta=1$ (blue dots) and
$\Delta=5/4$ (red dots) match each other and our prediction (dotted line), Eq.~(\ref{analy}), within
our numerical accuracy.}
\end{figure}

In general the exact phase diagram differs depending on our exact
choice for the dimension of $O$. 
However, the critical exponents
of the chiral transition do not. 
We refer the reader to Figs.~\ref{phases} and~\ref{nmfExp}, which illustrate these points.
This ``universality'' is one of our central
results. In fact, the critical exponents only depend upon the
dimension of the scalar field dual to the order parameter in the
effective AdS$_2$ region, $\Delta_{\rm IR}$. For example, near the
transition the order parameter scales as $\phi\sim
(O-O_c)^{\beta}$, where
\begin{equation}
\label{betaExp} \beta(\Delta_{\rm IR})=\left\{ \begin{array}{ll}
\frac{1}{2},  & \Delta_{\rm IR}
 \in \left[\frac{3}{4}, 1\right), \\ & \\
\frac{1-\Delta_{\rm IR}}{2\Delta_{\rm IR}-1}, & \Delta_{\rm IR}\in
\left(\frac{1}{2},\frac{3}{4}\right), \end{array}\right.
\end{equation}
In the D3/D5 system, $\Delta_{\rm IR}$ depends on the ratio of
magnetic field to density (\ref{dIR}),
\begin{equation}
\Delta_{\rm IR}=\frac{1 +
\sqrt{\frac{\tilde{d}^2-7B^2}{\tilde{d}^2+B^2}}}{2}, \nn
\end{equation}
where $B$ is the magnetic field and $\tilde{d}$ is proportional to the density. This dimension goes to unity at zero magnetic field and to $1/2$ at the holographic BKT transition, $B_c = \tilde{d}/\sqrt{7}$   (from where it
then enters the complex plane). 
%
This is a nice result: our line of
transitions not only exhibits non-mean-field exponents, but it
also continuously connects a transition with mean-field exponents at $\Delta_{\rm IR}=1$
to a holographic BKT transition at $\Delta_{\rm IR}=1/2$.
See also (\ref{analy}).

To understand this interpolation we describe the effective
potential  of the theory near the line of second-order
transitions. We do this numerically, finding that both the order
parameter and free energy in the broken phase follow from a
modified Landau-Ginzburg model with a potential of the form
\begin{equation}
\label{effLG}
V_{eff}(\phi)=\alpha_2 (O_c-O)\phi^2+\alpha_4 \phi^4+\alpha_{\rm IR} \phi^{1/(1-\Delta_{\rm IR})},
\end{equation}
where the $\alpha$'s are positive,  dimensionless couplings and
$\Delta_{\rm IR}$ is bigger than $1/2$ and less than $1$. When $O$
is tuned past its critical value $O_c$, the $\phi^4$ dominates for
$\Delta_{\rm IR}>3/4$ and the last term dominates for $\Delta_{\rm
IR}<3/4$. In the second regime, the static critical exponent
$\beta$ takes on a non mean-field value.  The non-analytic 
term in the
potential has a natural form given that there is an emergent $(0+1)$-dimensional infrared theory under which the condensate has dimension $1-\Delta_{\rm IR}$.
The term is just that required on dimensional
grounds in 
the infrared theory. We justify this further in Sec.~\ref{eff}.

Our thermodynamic results are obtained numerically. However, we do
obtain some analytic results for fluctuations in the symmetric
phase of the theory. Following~\cite{Faulkner:2009wj}, we perform
a matching computation to obtain the low-energy limit of the
two-point function of the order parameter. The result for the
brane system is essentially the same there. Assuming some basic
analyticity constraints, we compute both the correlation length
and the dynamical critical exponent near our transitions. The
former diverges with mean-field scaling while the latter assumes a
non mean-field value for all nonzero values of the magnetic field.
As with $\beta$, it only depends on $\Delta_{\rm IR}$.

What happens at nonzero temperature? In the previous examples with
both probe branes and holographic superfluids, the transitions are
universally second-order with mean-field
exponents~\cite{Jensen:2010ga}. Even BKT scaling is destroyed at
any nonzero temperature. For this reason, we also expect the
non-mean-field scaling of our transitions to be lost away from
zero temperature.  Indeed we find that this is the case below.

Our general conclusion then is that a conformal theory, perturbed
by  three control parameters $O_s$,$O_{1,2}$ with dimensions
$\Delta_s=\Delta_1>\Delta_2$ may lead to a phase diagram
qualitatively similar to that represented in Fig.~\ref{phases}. In
this general picture we require that $O_s$ tends to restore a
symmetry and $O_{1,2}$ to break it. The constraint that
$\Delta_s=\Delta_1$ amounts to the freedom to change the emergent
infrared theory in these systems. Meanwhile, the
$\Delta_1>\Delta_2$ condition allows us to trigger an ordinary
second-order transition without altering the infrared theory.
Since our results crucially depend upon an emergent CFT, we expect
that they extend beyond probe branes to all systems with such an
emergent theory.

We further test this picture phenomenologically by performing a
similar analysis of the D3/D7 system. In that theory, the magnetic
field has dimension two and the baryon density dimension three. We
show that if a chiral symmetry-breaking dimension three operator,
$O$, is also introduced, then the theory realizes a holographic
BKT transition at large $O$. As in the D3/D5 system, introducing a
control parameter with the same dimension as the density leads to
an emergent theory where the dimension of a scalar operator
depends on the density and $O$. Moreover, at intermediate $O$ the
magnetic field can be used to trigger a second-order
non-mean-field transition. While we have not computed it
numerically, we expect that the phase diagram qualitatively
matches our results for the D3/D5 system.

The outline of this work follows. In Sec.~\ref{setups} we review
the D3/D5 and D3/D7 systems at nonzero density and magnetic
field~\cite{Evans:2010iy,Jensen:2010vd,Jensen:2010ga,Evans:2010hi}. We go on to
present our phenomenological models and our holographic
regularization. The bulk of our results are presented in
Sec.~\ref{results}, beginning with the thermodynamics of the D3/D5
system at zero and nonzero temperature. Next, we study
fluctuations using a matching procedure. In Sec.\ref{eff} we
present our effective theory of the transition and critically test
it. We apply our analysis briefly to the D3/D7 system without
numerics in Sec.\ref{d3d7}. Finally, we discuss our results in
Sec.~\ref{discuss}.

\section{Holographic setups}
\label{setups}
\subsection{The D3/D5 system}
Strongly coupled $SU(N)$ ${\cal N}$=4 super Yang-Mills  (SYM)
theory at large $N$ and zero temperature is dual to type IIB
supergravity on AdS$_5 \times S^5$ (with $N$ D3 branes at its
core). The geometry can be written as
\begin{equation}
   ds^2 =\frac{w^2}{R^2}dx^2_{3,1}+\frac{R^2}{w^2}(d\rho^2+\rho^2d\Omega_2^2+dL^2+L^2d\bar{\Omega}_2^2),
\end{equation}
where $w^2=\rho^2+L^2$, $d\Omega_2^2$, $d\bar{\Omega}_2^2$ are
the metrics for two unit two-spheres, and $R^4=4 \pi g_s N
\alpha^{'2}$. In these coordinates the Poinc\'are horizon of AdS
is located at $\rho=L=0$ and the boundary as
$\rho^2+L^2\rightarrow\infty$.

We now add $N_f$ flavor hypermultiplets to the gauge theory along
a $(2+1)$  dimensional defect by placing a probe D5 brane in this
geometry. The probe limit corresponds to the quenched limit of the
gauge theory. The D5 probes are described by their DBI action
\beq S_{DBI} = - N_fT_{5} \int d^6\xi \sqrt{- {\rm det} (P[G]_{ab} +
F_{ab})} \ , \eeq
where $a,b=0,..,5$ are worldvolume indices, $P[G]_{ab}$ is the
pullback  of the metric to the brane, and $F$ is the field
strength for the diagonal $U(1)$ gauge field living on the D5
worldvolume. The field theory has a $SO(3)_1\times SO(3)_2\times
U(1)_B$ global symmetry, where the two $SO(3)$'s are chiral
R-symmetries and the $U(1)_B$ is a baryon number symmetry which
only rotates the flavor fields. The baryon symmetry current is
dual to the $U(1)$ gauge field on the brane while the chiral
symmetries correspond to the rotational symmetry of the two
two-spheres. We now consider an ansatz wherein our brane
embeddings are translationally invariant in the wrapped field
theory directions $x^0-x^2$ while wrapping the first two-sphere
and the ``radial coordinate'' $\rho$. We also consider the theory
with no source for the baryon current, that is with $F=0$. There
are then a family of embeddings $L=m$ and $x^3,\theta_2,\phi_2$
constant (where $\theta_2,\phi_2$ are coordinates on the unwrapped
two-sphere). These are supersymmetric embeddings that correspond
to the theory with a hypermultiplet of mass $m$.  The second
$SO(3)$ chiral symmetry is then explicitly broken by a hyper mass
and spontaneously broken by a vev for the corresponding operator
$\bar{\psi}\psi$ (plus operators related by supersymmetry) 
~\cite{Kruczenski:2003be}. It is this
chiral symmetry that is spontaneously broken in our chiral
transitions.

We now extend our ansatz to include nonzero  baryon density and
magnetic field. To do this, we need to let the embedding function
$L$ depend on $\rho$ as $L=L(\rho)$ as well as turn on a
non-trivial field strength~\cite{Karch:2007br,Filev:2007gb},
\begin{equation}
F=A_0'(\rho)d\rho\wedge dx^0 + B\, dx^1\wedge dx^2,
\end{equation}
where the field $A_0$ will determine both  the chemical potential
and density of baryon charge \cite{Kim:2006gp,Kim:2007zm,Nakamura:2006xk,Kobayashi:2006sb} and $B$ is the magnetic field. The
radial electric field $A_0'$ is sourced by charge at the bottom of
AdS, and so the brane ends there. This amounts to the boundary
condition $L(0)=0$. For this ansatz we can consistently neglect
the Wess-Zumino pieces of the brane action and write
\begin{align}
  \nonumber
  \label{S5}
  S_5 =-&\underbrace{N_fT_5R^6 \text{vol}[S^2]}_{\equiv \mathcal{N}}\text{vol}
  [\mathbb{R}^{2,1}] \\ &\times\int d\rho \,\rho^2\sqrt{1+L'^2-A_0'^2}\sqrt{1+\frac{B^2}{w^4}},
\end{align}
where we have defined $w^2=\rho^2+L^2$ as well as rescaled  $x$
and  $\rho$ by powers of $R$. The normalization is given by
$\mathcal{N}=\sqrt{\lambda}N/2\pi^3$ where $\lambda=4\pi g_{\rm
YM}^2N$ is the 't Hooft coupling of the SYM theory. From here
onward, we will refer to the action \emph{density}
$S_5/\text{vol}[\mathbb{R}^{2,1}]$, describing it with the same
notation $S_5$.

Notably, the action only depends on $A_0$ through its radial
derivative.  Thus there is a conserved quantity,
\begin{equation}
d= \frac{\delta S_5}{\delta A_0'(\rho,x)}=\frac{\mathcal{N}\rho^2A_0'\sqrt{1+\frac{B^2}{w^4}}}{\sqrt{1+L'^2-A_0'^2}}.
\end{equation}
In fact, $d$ is the baryon density in the dual theory.  Solving
for $A_0'$ in terms of a rescaled density
$\tilde{d}=d/\mathcal{N}$, we find
\begin{equation}
A_0'^2=\frac{\tilde{d}^2(1+L'^2)}{\tilde{d}^2+\rho^4\left( 1+\frac{B^2}{w^4}\right)}.
\end{equation}
We obtain the brane action at  fixed density by substituting this
result into the action Eq.~\ref{S5} and Legendre transforming with
respect to $A_0'$. The result is
\begin{equation}
\label{newS5}
\tilde{S}_5=-\mathcal{N}\int d\rho\sqrt{1+L'^2}\sqrt{\tilde{d}^2+\rho^4\left(1+\frac{B^2}{w^4}\right)}.
\end{equation}
Field configurations $L(\rho)$ that extremize this action
correspond to field theory ensembles that extremize the effective
potential of the theory in the canonical ensemble. In general,
these configurations can only be obtained numerically. However
there is an exact solution to the equation of motion for all
$\tilde{d}$ and $B$, $L=0$, which corresponds to the dual theory
with zero hyper mass and zero condensate. This solution
corresponds to the chirally symmetric phase of the theory.

We continue by reviewing the origin  of the chiral BKT transition
in this system. The onset of the transition can be understood by
studying the stability of the symmetric embedding. Small
fluctuations around $L=0$ are described by the quadratic piece of
Eq.~(\ref{newS5}),
\begin{equation}
\tilde{\mathcal{L}}_5\sim -\frac{\mathcal{N}}{2}
\sqrt{\tilde{d}^2+B^2+\rho^4}L'^2+\frac{\mathcal{N}B^2L^2}{\rho^2\sqrt{\tilde{d}^2+B^2+\rho^4}}.
\end{equation}
This Lagrangian has two distinct limits \footnote{In our analysis
we use the results for a scalar in AdS$_{p+1}$: The solution of
the equation of motion is
\begin{eqnarray}
  && \frac{L}{\rho} \sim \left(\frac{1}{\rho}\right)^\D  \\
  && \Delta_{\pm} = {p \over
2} \pm \sqrt{\left( {p \over 2} \right)^2 + m^2}  \ .
  \label{Delta}
\end{eqnarray}
and the Breitenlohner-Freedman (BF)
bound~\cite{Breitenlohner:1982jf} is given by $-p^2/4$}. At large
$\rho\gg \sqrt{B},\sqrt{\tilde{d}}$, the field $L/\rho$ fluctuates
as a stable $m^2=-2$ scalar field in AdS$_4$. However, at small
$\rho\ll \sqrt{B},\sqrt{\tilde{d}}$, $L/\rho$ fluctuates as a
$m^2=-2B^2/(\tilde{d}^2+B^2)$ scalar in AdS$_2$. Thus for
$\tilde{d}/B> \sqrt{7}$ the field is stable but for
$\tilde{d}/B<\sqrt{7}$ the mass drops below the
Breitenlohner-Freedman (BF) bound~\cite{Breitenlohner:1982jf} in
AdS$_2$, $m_{\rm BF}^2=-1/4$. There is therefore a chiral
transition at the critical filling fraction
\begin{equation}
\nu_c=\frac{d}{B}_c=\frac{\sqrt{7\lambda}N_fN}{2\pi^3}.
\end{equation}
Further analysis reveals that the order parameter  scales
exponentially at smaller densities, so that the transition is of
the holographic BKT type. This behaviour is the result of the
violation of the BF bound in the infrared region which implies
that an infinite number of tachyons form at the transition, an
extremely unnatural situation within the Landau-Ginzburg-Wilson
paradigm. Because it will be important in the rest of this paper,
we also note that in the effective AdS$_2$ region $L/\rho$ is dual
to a scalar operator in the emergent CFT of dimension
\begin{equation} \label{dIR}
\Delta_{\rm IR}=\frac{1 +
\sqrt{\frac{\tilde{d}^2-7B^2}{\tilde{d}^2+B^2}}}{2},
\end{equation}
As usual in AdS/CFT there are two solutions to the equation
of motion for the scalar in AdS (see Eq.~(\ref{Delta})) which describe
an operator in the field theory and its source. The second
solution here corresponds to an object of dimension of $1-\Delta_{IR}$. We will see
below when we discuss the effective theory for our transitions
that the 3d condensate corresponds to a dimension $1-\Delta_{\rm IR}$ operator in the infrared theory.

At zero magnetic field we have $\Delta_{\rm IR}=1$, which
decreases to $\Delta_{\rm IR}=1/2$ at the transition. In the
broken phase, $\Delta_{\rm IR}$ is driven into the complex plane.
In this way the scaling symmetry of the infrared theory is broken
to a discrete subgroup (which is broken further to a self-similar
subset by higher energy physics), which relates the various
tachyons of the symmetric phase.

\subsection{The D3/D7 system}
In the same way we can consider strongly coupled $\mathcal{N}=4$
SYM at large $N$ coupled to $(3+1)$ dimensional fundamental
hypermultiplets. In the quenched limit the flavor fields are well
described by probe D7 branes in the AdS$_5\times S^5$ geometry.
The global symmetry of this theory is $SO(4)\times
U(1)_{\chi}\times U(1)_B$, where the $U(1)_{\chi}$ is a chiral
symmetry and the $U(1)_B$ is the usual baryon number symmetry.
This chiral symmetry is explicitly broken by a hyper mass and
spontaneously broken by a condensate of the hyper mass operator.

On the gravity side, the D7 branes wrap a three-sphere rather
than a two-sphere. The $U(1)_{\chi}$ chiral symmetry is dual to
the $SO(2)$ isometry of an $\mathbb{R}^2$ transverse to both
stacks of D3 and D7 branes. The baryon symmetry current is dual to
the diagonal $U(1)$ gauge field on the D7 branes as before. We
then consider a translationally-invariant, $SO(4)$-preserving
ansatz as before with a density and magnetic field. For such an
ansatz the brane action at fixed density is
\begin{equation}
\label{S7}
\tilde{S}_7=-\mathcal{N}_7\int d\rho\,\sqrt{1+L'^2}\sqrt{\tilde{d}^2
+\rho^6\left(1+\frac{B^2}{w^4}\right)}.
\end{equation}
As above, we can study the onset of the chiral  transition by
studying the stability of small fluctuations around the chirally
symmetric embedding $L=0$. These are described by the quadratic
part of Eq.~(\ref{S7}),
\begin{equation}
\tilde{\mathcal{L}}_7\sim -\frac{\mathcal{N}_7}{2}\sqrt{\tilde{d}^2+\rho^2B^2
+\rho^6}L'^2+\frac{\mathcal{N}_7B^2L^2}{\sqrt{\tilde{d}^2+\rho^2B^2+\rho^6}}.
\end{equation}
As with the D3/D5 system, this Lagrangian has two  distinct
limits. For $\rho\gg \tilde{d}^{1/3},\sqrt{B}$, $L/\rho$
fluctuates as a stable $m^2=-3$ scalar field in AdS$_5$. On the
other hand, for $\rho\ll \tilde{d}^{1/3},\sqrt{B}$ $L/\rho$
fluctuates as a stable massless scalar in AdS$_2$. As originally
pointed out in~\cite{Jensen:2010vx}, there is an emergent CFT in
this theory as well. Moreover, $L/\rho$ is dual to a scalar
operator in the infrared theory of dimension $\Delta_{\rm IR}=1$.

There is no holographic BKT transition in this system.  Rather,
the chiral transition is second-order with mean-field exponents. A
single tachyon forms at the transition, which is effectively
modelled by a Landau-Ginzburg model with a quartic potential.
Later, we will see that the mean-field exponents are crucially
related to the fact that $\Delta_{\rm IR}=1$. For the
majority of this paper we will work in the D3/D5 system that does
have a BKT transition but we will return at the end to produce
similar phenomena in a phenomenological deformed version of this
D3/D7 system.

\subsection{Phenomenological models}
\label{pheno} We seek to extend the brane systems above by turning
on a third control parameter. For computational simplicity, we
seek to deform our setups in such a way that the brane action
depends only upon a single worldvolume field and a number of
constants of the motion. For the D3/D5 system, there are several
candidates. The first is an electric field along the
brane~\cite{Karch:2007pd} and the second a flux on the wrapped
two-sphere~\cite{Myers:2008me}. Neither deformation breaks chiral symmetry 
at zero magnetic field and zero hyper mass, but the electric field may yet lead to interesting results.
Other deformations involve additional worldvolume fields
whose equations of motion are not integrable.

In favor of solving a more complicated brane  problem with at
least two worldvolume fields, we elect to take a phenomenological
approach. We will simulate a chiral symmetry-breaking control
parameter whose dimension we dial. At first, this approach may
seem cavalier: in contrast with the ``bottom-up'' holographic
superfluid and non-Fermi liquid analyses, there are many different
ways that control parameters emerge in a probe brane action. There
are few \emph{a priori} reasons to believe that phenomenology will
accurately predict features of transitions in consistent
``top-down'' brane setups with three control parameters.

The best justification for our method comes  \emph{ex post facto}.
Ultimately, we find that the critical exponents we measure do not
depend upon the details of our simulated deformation. This result
is crucial and we will return to it extensively later. For now, we
will simply describe our phenomenological choice. We simulate a
control parameter $O$ of dimension $\Delta$ (taken to be relevant)
in the D3/D5 system by considering a modified brane action
\begin{equation}
\label{phenoS5}
\tilde{S}_5=-\mathcal{N}\int d\rho\,\sqrt{1+L'^2}
\sqrt{\tilde{d}^2+\rho^4\left( 1+\frac{B^2}{w^4}+\frac{O^2}{w^{2\Delta}} \right)}.
\end{equation}
Note that when $O$ has dimension two, it is effectively a magnetic field.

Earlier we studied both the onset of the chiral transition as well
as the emergent infrared theory by studying small fluctuations
around the symmetric $L=0$ embedding. These are now described by
the quadratic Lagrangian,
\begin{align}
\nonumber
\tilde{\mathcal{L}}_5\sim &-\frac{\mathcal{N}}{2}
\sqrt{\tilde{d}^2+B^2+\rho^{4-2\Delta}O^2+\rho^4}L'^2 \\
&+\left( \frac{B^2}{\rho^2}+\frac{O^2\Delta}{2\rho^{2(\Delta-1)}} \right)
\frac{\mathcal{N}L^2}{\sqrt{\tilde{d}^2+B^2+\rho^{4-2\Delta}O^2+\rho^4}}.
\end{align}
This system has two different infrared limits depending upon the
value  of $\Delta$. For $\Delta>2$, the new control parameter
dominates the infrared, so that $L/\rho$ fluctuates as an unstable
scalar there. Then for any nonzero $O$ the symmetric embedding is
unstable and the stable phase will break chiral symmetry. For
$\Delta<2$ the magnetic field and density together dominate the
infrared. As before, at small $\rho\ll \sqrt{B},\sqrt{\tilde{d}}$,
the field $L/\rho$ fluctuates as a $m^2=-2B^2/(\tilde{d}^2+B^2)$
scalar in an effective AdS$_2$ region at the bottom of the brane.
Finally, the introduction of $O$ tends to break chiral symmetry.
We see this by studying small fluctuations at nonzero $O$ but
vanishing density and magnetic field. In this limit, the radial
equation of motion for the field $L/\rho$ at small $\rho$ is that
of a $m^2=\Delta-3$ scalar in AdS$_{4-\Delta}$. Provided that $O$
is not marginal with $\Delta=3$, $L/\rho$ fluctuates unstably in
the infrared and so the true ground state breaks chiral symmetry
as claimed.

We can make a similar phenomenological deformation to introduce an
operator $O$ of arbitrary dimension into the D3/D7 system. Just as
in the D3/D5 system, we introduce a simulated control parameter
into the brane action as
\begin{equation}
\label{phenoS7} \tilde{S}_7=-\mathcal{N}_7\int
d\rho\,\sqrt{1+L'^2} \sqrt{\tilde{d}^2+\rho^6\left(
1+\frac{B^2}{w^4}+\frac{O^2}{w^{2\Delta}} \right)}.
\end{equation}
We will discuss the physics of this model in the later Sec.\ref{d3d7}.

\subsection{Holographic regularization}
The phenomenological brane actions
Eq.~(\ref{phenoS5}),(\ref{phenoS7})  contain a number of
near-boundary divergences~\cite{Karch:2005ms}. In a genuine
``top-down'' construction, these correspond to ultraviolet
divergences of the dual theory. In the bulk, they can be
diffeomorphism-invariantly regulated by introducting a
near-boundary cutoff slice and adding local counterterms on the
slice. This process is known as holographic renormalization and is
crucial: once we have an appropriately renormalized bulk action,
we may sensibly take derivatives to obtain correlation functions.

In our phenomenological constructions, however, we cannot  be sure
that we diffeomorphism-invariantly regulate the bulk theory.
Rather we choose to regularize our theories in a manner inspired
by holographic renormalization. We introduce a cutoff slice, add
counterterms, and then take the cutoff to infinity. To illustrate
the idea, we consider the modified D3/D5 system of
Eq.~(\ref{phenoS5}), where $O$ has dimension $\Delta=1$. A general
solution $L(\rho)$ has the near-boundary solution
\begin{equation}
\label{bdyLD5}
L(\rho)=m+\sum_{n=1}^{\infty}\frac{L_n}{\rho^n},
\end{equation}
where the $L_n$ for $n>1$ are recursively  determined by $m$ and
$L_1$. The parameter $m$ is simply the hyper mass. Then the brane
action, integrated up to a cutoff $\rho=\Lambda$, evaluated on
such a solution has the near-boundary divergence structure
\begin{equation}
\tilde{S}_{5,\Lambda}=-\mathcal{N}\int d^3x\left(\frac{\Lambda^3}{3}+\frac{O^2\Lambda}{2}
 \right)+\text{finite}.
\end{equation}
The exact divergences depend upon $\Delta$; for $\Delta>3/4$ there
is only a single counterterm required, while for $\Delta=3/(2n)$
for $n$ a positive integer there is also a logarithmic divergence.
These logarithmic divergences correspond to Weyl anomalies of the
dual theory~\cite{Henningson:1998gx}. In this case of $\Delta=1$,
we add simple counterterms on the cutoff slice,
\begin{equation}
\tilde{S}_{\rm CT}=\mathcal{N}\int_{\rho=\Lambda} d^3x\sqrt{-\gamma}\left(\frac{1}{3}+\frac{O^2}{2\rho^2}\right),
\end{equation}
where $\gamma$ is the induced metric on  the slice. When there are
logarithmic divergences, we subtract them with counterterms of the
form $\sqrt{-\gamma}O^{j}/(\rho^{j}\log \rho)$. In general, we
define a regularized action by
\begin{equation}
\tilde{S}_{5,\rm reg}\equiv\lim_{\Lambda\rightarrow\infty}\left[ \tilde{S}_{5,\Lambda}+S_{\rm CT}\right].
\end{equation}
We now \emph{define} correlation functions of  the dual theory
through functional derivatives.

For $\Delta>1/2$ there are no divergences that  depend on $m$ or
$L_1$. Consequently in this region the one-point function of the
operator dual to $L$, the hyper mass operator $\mathcal{O}_L$, is
simply
\begin{equation}
\phi=\langle \mathcal{O}_L\rangle = -\frac{\delta \tilde{S}_{5,\rm reg}}{\delta m}=-\mathcal{N}L_1.
\end{equation}
On the other hand, for $\Delta\leq 1/2$  there are additional
contributions to $\langle \mathcal{O}_y\rangle$ that arise from
extra $L$-dependent counterterms. These counterterms are
proportional to $L(\rho)^2$ and so lead to a contribution to the
condensate proportional to $m$. These contributions vanish in the
chiral limit we consider.

A similar analysis can be performed for the  D3/D7 system. Since
we do not use the results in this work, we simply quote the
highlights. For the choice $\Delta=3$ that we consider in this
work, there are only two divergences: the first corresponds to the
infinite volume on the cutoff slice and the second to a
logarithmic divergence proportional to $F_{\mu\nu}F^{\mu\nu}$
evaulated on the slice. The regularization is then identical to
the holographic renormalization performed in~\cite{Jensen:2010vd}.
Notably, the logarithmic divergence corresponds to a Weyl anomaly
of the dual theory: the trace of the stress tensor of the dual
theory is proportional to $F^2$.

\section{Results}
\label{results} In this section we report the bulk of our
numerical results. We begin by studying the zero temperature
transitions of our deformed D3/D5 system. As a first numerical
check we reproduce the BKT transition in the D3/D5 system with a
magnetic field and density. Next, we study this system with the
control parameter $O$ present but at zero magnetic field and show
that it triggers a mean-field chiral transition. Next, we study
the zero temperature phase diagram in the $(O,B)$ plane and
identify the line of quantum critical points. We go on to study
the nonzero temperature transitions, for which the non mean-field
scaling is destroyed. Finally we compute the low-energy behavior
of the two-point function of the order parameter in the symmetric
phase.

\subsection{Zero temperature transitions}

\subsubsection{BKT transition} \label{basnum}

We begin by studying the D3/D5 system with magnetic field and
density to reproduce the BKT transition~\cite{Jensen:2010ga}. 
To do so we must solve
the equation of motion for $L(\rho)$ that follows from varying
Eq.~(\ref{phenoS5}) with $O$ zero. This equation can 
be solved numerically. We do so with the shooting method, generally
shooting from large $\rho$.

Recall that the radial electric field on the brane is sourced by
charge at the bottom of the geometry. This is equivalent to
setting the infrared boundary condition $L(0)=0$. At small $\rho$
there is then a series solution for $L$
\begin{equation}
L(\rho)=\gamma_0+\gamma_1\rho+\sum_{n=2}^{\infty}\gamma_n \rho^n,
\end{equation}
where the higher $\gamma_n$ are recursively  determined by
$\gamma_0$ and $\gamma_1$.

Near the AdS$_4$ boundary we impose the boundary condition that
our dual flavor is massless. This amounts to choosing the leading
term in the near-boundary solution Eq.~(\ref{bdyLD5}) to vanish.
For our shooting we use the large $\rho$ series solution as
initial data (having computed the first dozen or so $L_n s$) upon
which we numerically integrate the corresponding solution to small
$\rho$. We then match this solution onto the small $\rho$
solution. By dialing the field theory condensate we shoot for
solutions that extend to the bottom of AdS with $\gamma_0=0$. It
is numerically difficult to match onto a small $\rho$ series
solution. A similar problem emerges in the small $\rho$ embeddings
in the flavored little string theory studied
in~\cite{Jensen:2010vx}: in each case, there is a non-analyticity
in the equation of motion for $L$ at $\rho=0$. For the D3/D5
system, embeddings consequently ``spike'' to $\rho=0$, infinitely
so at infinitesimally small $\rho$. Nonetheless, with care, we
have managed to successfully shoot to the infrared boundary
condition with high accuracy.

\begin{figure}[]
\centering
  \subfigure[{\small The embedding $L$ of a D5 brane in the D3 geometry
  for various $B/ \tilde{d}$ showing the BKT transition.  }]
  {\includegraphics[width=7cm]{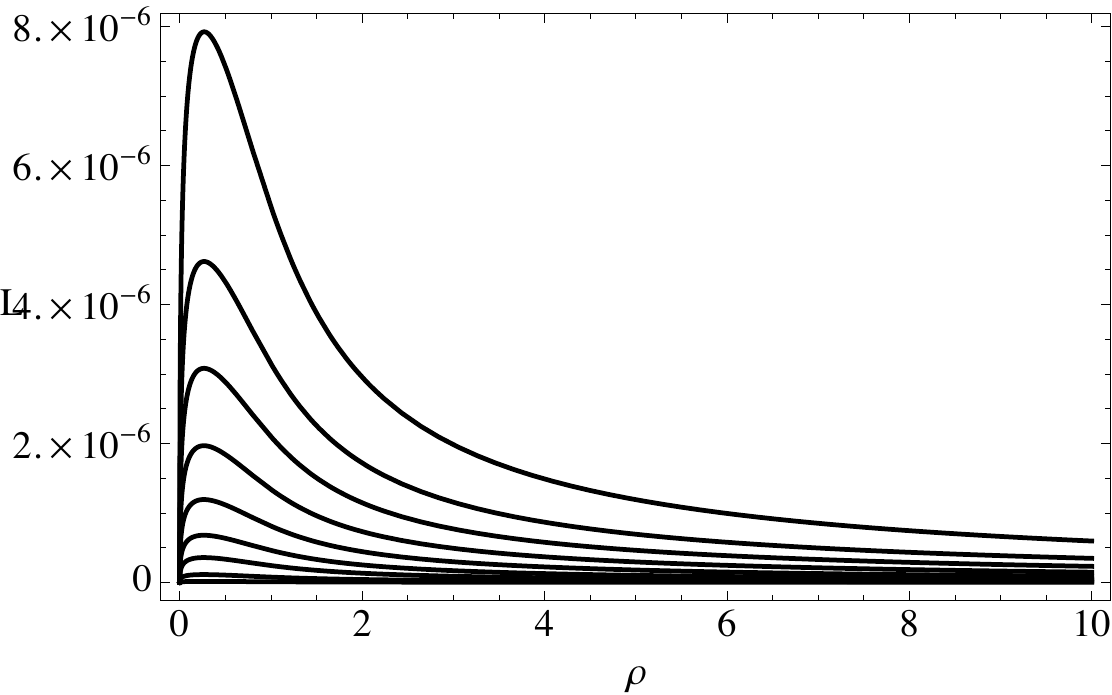}}
  \subfigure[{\small A plot of the quark condensate $c$ versus $B$
  across the D3/D5 BKT transition.
 }]
  {\includegraphics[width=7cm]{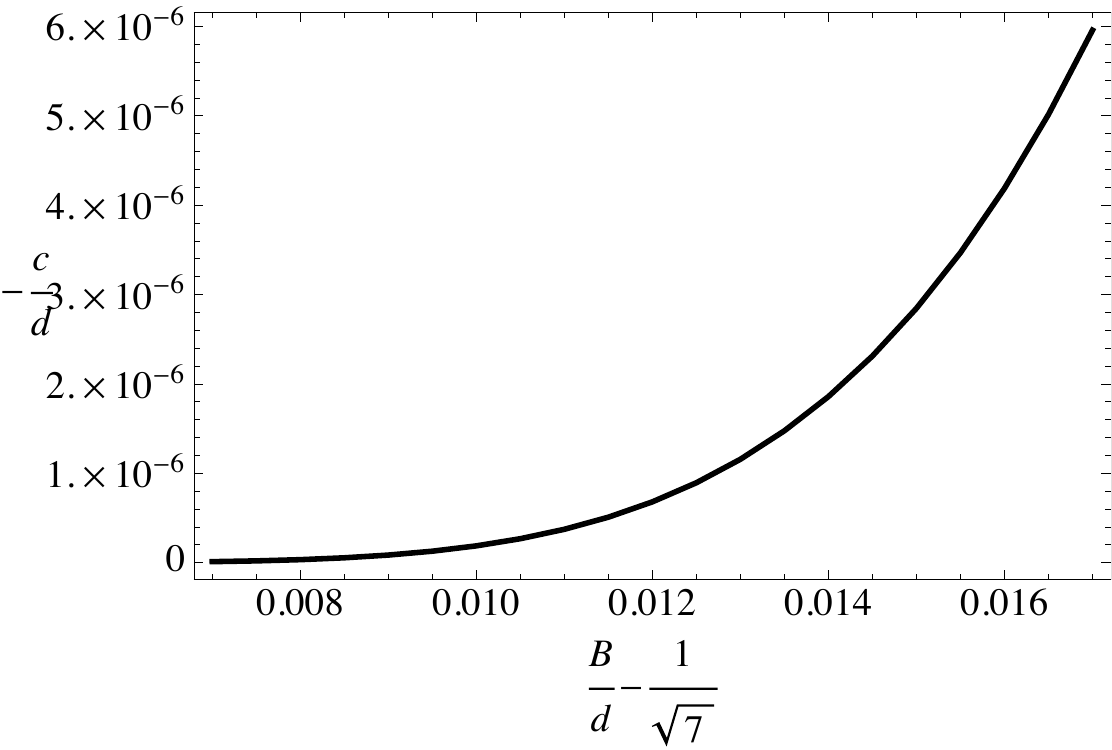}}
  \caption{
           {\small The BKT transition in the D3/D5 system with quark density
           and magnetic field present. }
           }\label{BKT}
\end{figure}

Generically with large $B$ the solutions bend off the $L=0$ axis.
The chemical potential then forces the solution to spike to the
origin at $\rho=L=0$. At some critical value of $B$ the embeddings
smoothly transition to the $L=0$ embedding. These embeddings are
shown in Fig \ref{BKT} for varying $B / \tilde{d}$. We also plot
the value of the quark condensate, $c$, across the embedding to
show the BKT exponential scaling.

Note that at yet larger $B$ there is a second transition to a
phase with $\tilde{d}=0$ and stable mesons described by embeddings
that curve off the axis but do not spike to the origin. That
transition is described for this theory in \cite{Evans:2010hi} but
we will not explore it further in this paper.

\subsubsection{The phenomenological operator O}
Next we study the deformed D3/D5 system at zero magnetic field. As
we discussed in Sec.~\ref{pheno} when the dimension of $O$ is
greater than two we find that the chirally broken phase is
preferred for all values of the density. This matches the
prediction above that the $L=0$ embedding is unstable for all
$\tilde{d}$.

When the dimension of $O$ is less than two, we expect large $O$ at
fixed density to trigger a chiral transition. We seek to locate
this transition for many different dimensions $\Delta$ and to measure the
associated critical exponents. To do so we must solve the equation
of motion for $L(\rho)$ that follows from varying
Eq.~(\ref{phenoS5}) with $B=0$. Again we use the numerical
techniques discussed in Sec.~\ref{basnum} above.

We find that for large $O\gg \tilde{d}^{\Delta/2}$,  the solutions
corresponding to zero mass bend away from the symmetric embedding
$L=0$. Near the AdS$_4$ boundary they necessarily asymptote to the
symmetric embedding, but at small $\rho$ they ``spike'' to the
bottom of AdS. Thus chiral symmetry is indeed broken at large $O$
as expected.

\begin{figure}[]
\includegraphics[width=8cm]{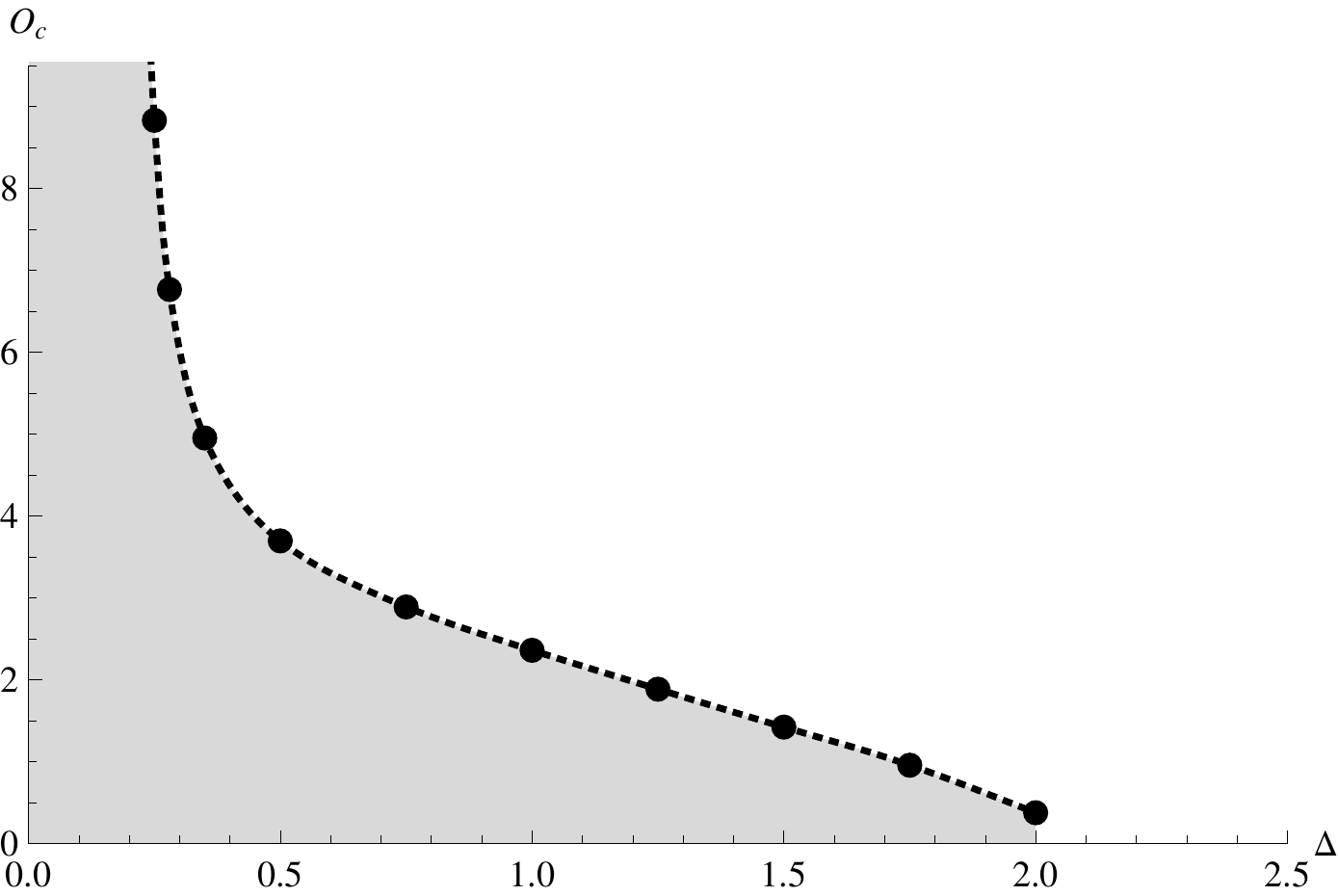}
\caption{\label{meanField}The zero temperature ``phase diagram''
of the deformed D3/D5 theory at zero magnetic field as a function
of the dimension and value of the deformation $O$. The shaded
region is the chirally symmetric phase and the white is the broken
phase. The line of transitions is second-order with mean-field
exponents, excepting a holographic BKT transition at $\Delta=2$.}
\end{figure}

\begin{figure}[]
\includegraphics[width=8cm]{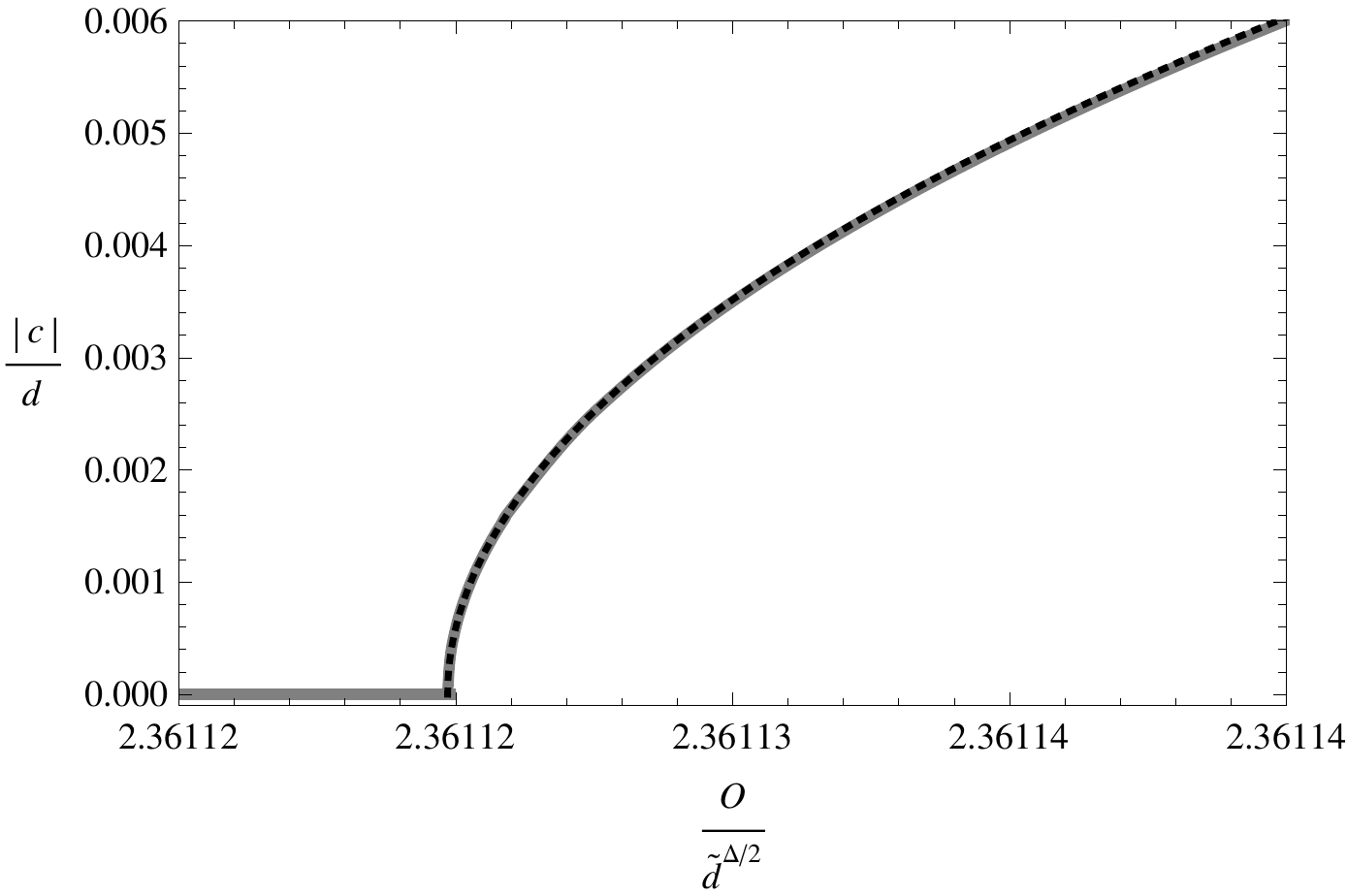}
\caption{\label{mfCond} The condensate in the deformed D3/D5
theory at zero magnetic field and $\Delta=1$. The solid line is
numerical data and the dotted a fit. Near the transition the
condensate scales with a mean-field exponent, $c\sim
\sqrt{O-O_c}$.}
\end{figure}

For fixed $\Delta \leq 2$, there is a critical $O_c$ where the
embeddings smoothly transition to the symmetric embedding. This is
the location of the chiral transition. We locate the transition
for many different $\Delta$, for which we must also scan through
$O$. This is somewhat laborious, as we have to shoot for each
value of $O$ and $\Delta$. The net result is shown in
Fig.~\ref{meanField}. For $\Delta<2$ we identify a second-order
transition with mean-field exponents as promised. 
At $\Delta=2$ there is a holographic BKT transition, as $O$ then acts like a magnetic field.
We plot the
condensate near the transition for the particular case $\Delta=1$
in Fig.~\ref{mfCond} to display that mean field behavior.

\subsubsection{The general case}
\label{nmfSec} Now we turn on the magnetic field and $O$ together.
This will tune the dimension $\Delta_{IR}$ of the operator dual to $L/\rho$ in
the emergent theory as we discussed in Sec.~\ref{pheno}. As above,
we can only solve the equation of motion for $L$ numerically. Our
procedure is essentially the same as the one described above in
Sec.~\ref{basnum}.

We studied two different dimensions $\Delta$ for $O$  in great
detail, $\Delta=1$ and $\Delta=5/4$. The choice of a
non-integer dimension explcitly shows the independence
of our results on that dimension. In each case, we studied the
chiral transition at many different magnetic fields and
consequently at many values of $O$ as well. For each such choice
of $O$ and $B$ we must shoot to find the correct vacuum. As
expected, we find a line of second-order chiral transitions. The
resulting phase diagram as a function of $O$ and magnetic field
was already plotted in Fig.~\ref{phases}. We also plot the
critical exponent $\beta$ as a function of magnetic field in
Fig.~\ref{nmfExp}. Recall that $\beta$ is the scaling of the order
parameter in the broken phase, $c\sim (O-O_c)^{\beta}$.

The combined results are very interesting. First, the line of
transitions is second-order as expected. Moreover, for small
magnetic fields such that $\Delta_{\rm IR}>3/4$, the exponent
$\beta$ takes on the mean-field value $1/2$. Once the infrared
dimension dips below $3/4$, however, the exponent $\beta$ is no
longer $1/2$ and moreover is independent of the dimension of $O$.
The simplest way to interpret this result is that the effective
potential of the theory near the transition has the usual quartic
term as well as a second term that depends upon $\Delta_{\rm IR}$
but not $\Delta$. We will show how this occurs explicitly in
Sec.~\ref{eff}, where we construct a modified Landau-Ginzburg
model for the transition.

\subsection{Nonzero temperature thermodynamics}

It is interesting to also study the behaviour of our model at
nonzero temperature. At nonzero temperature, the $\mathcal{N}=4$
SYM  theory is holographically described by IIB supergravity on an
AdS$_5$ black brane geometry (with $N$ hot D3 branes at its core).
The geometry can be written as
\begin{equation}
   ds^2 =\frac{w^2}{R^2}(-f(w)(dx^0)^2+d\vec{x}^2)+\frac{R^2}{f(w)w^2}dw^2+R^2d\Omega_5^2
\end{equation}
where
\begin{equation}
f(w)=1-\frac{w_h^4}{w^2}, \,\,\, d\Omega_5^2 = d\theta^2+\cos^2
\theta d\Omega_2^2+\sin^2\theta d\bar{\Omega}_2^2
\end{equation}
and we define $w_h=\pi T/R^2$ with $T$ the temperature of both the
field and gravitational theories. This coordinate system is
related to the one we employ at zero temperature by
$L=w\sin\theta,$ $\rho=w\cos\theta$. Both $\theta$ and $L$ are
dual to the hyper mass operator. We change coordinates simply
because we have found the numerics easier for this analysis.

As before, we embed $N_f$ D5 branes in this geometry. We consider
embeddings that are translationally-invariant in the wrapped
$x^0-x^2$ directions, wrap the first two-sphere and $w$, and
possess no angular momentum on either two-sphere. The embedding is
parametrized by the worldvolume field $\theta=\theta(w)$. After
adding a charge density and magnetic field, the D5 action at fixed
density is
\begin{align} \label{Stemp}
\tilde{S}_5&=-\mathcal{N}\int dw \sqrt{1+fw^2\theta'^2}
\sqrt{\tilde{d}^2+w^4\cos^4\theta\left(1+\frac{B^2}{w^2}\right)}.
\end{align}
Now we add our third control parameter $O$. There is yet further
ambiguity in how we phenomenologically introduce $O$. We elect to
add it in such a way that it again becomes a contribution to the
magnetic field when the dimension of $O$ approaches two. Our
deformed Lagrangian is
\begin{equation}
\label{Ltemp}
\tilde{\mathcal{L}}_5=-\mathcal{N}\sqrt{1+fw^2\theta'^2}
\sqrt{\tilde{d}^2+w^4\cos^4\theta\left(1+\frac{B^2}{w^4}+\frac{O^2}{w^{2\Delta}}\right)}.
\end{equation}

At zero temprature the holographic BKT Transition was triggered by
driving  the mass of $L$ in the effective AdS$_2$ region below the
BF bound. At any nonzero temperature this exponential scaling is
lost~\cite{Jensen:2010ga,Evans:2010hi}. The infrared region
becomes an AdS$_2$-like space with a black hole, which has a
Rindler near-horizon limit. Driving the mass below the BF bound
then corresponds to a UV instability (but not an IR instability)
from the point of view of the infrared theory. This instability is
tamed by the ultraviolet completion to AdS$_4$ physics. The
absence of such an IR instability presumably leads to the
resulting second-order mean-field transition observed at extremely
small temperatures in the D3/D5 system.

By the same logic we expect the non mean-field scaling of our
second-order  transitions to be destroyed at any nonzero
temperature. Indeed, we find this result numerically. To do this,
we  extremize the modified action Eq.~(\ref{Ltemp}) and regularize
the bulk action in such a way that we measure the field theory
condensate from our embeddings. As at zero temperature, we employ
a shooting technique. This time we elect to shoot from the
infrared near the black brane horizon. The charge on the brane
indicates that the embedding extends down to the horizon. Our
infrared boundary condition is then simply that the embedding is
regular there.

\begin{figure}[]
  \includegraphics[width=8cm]{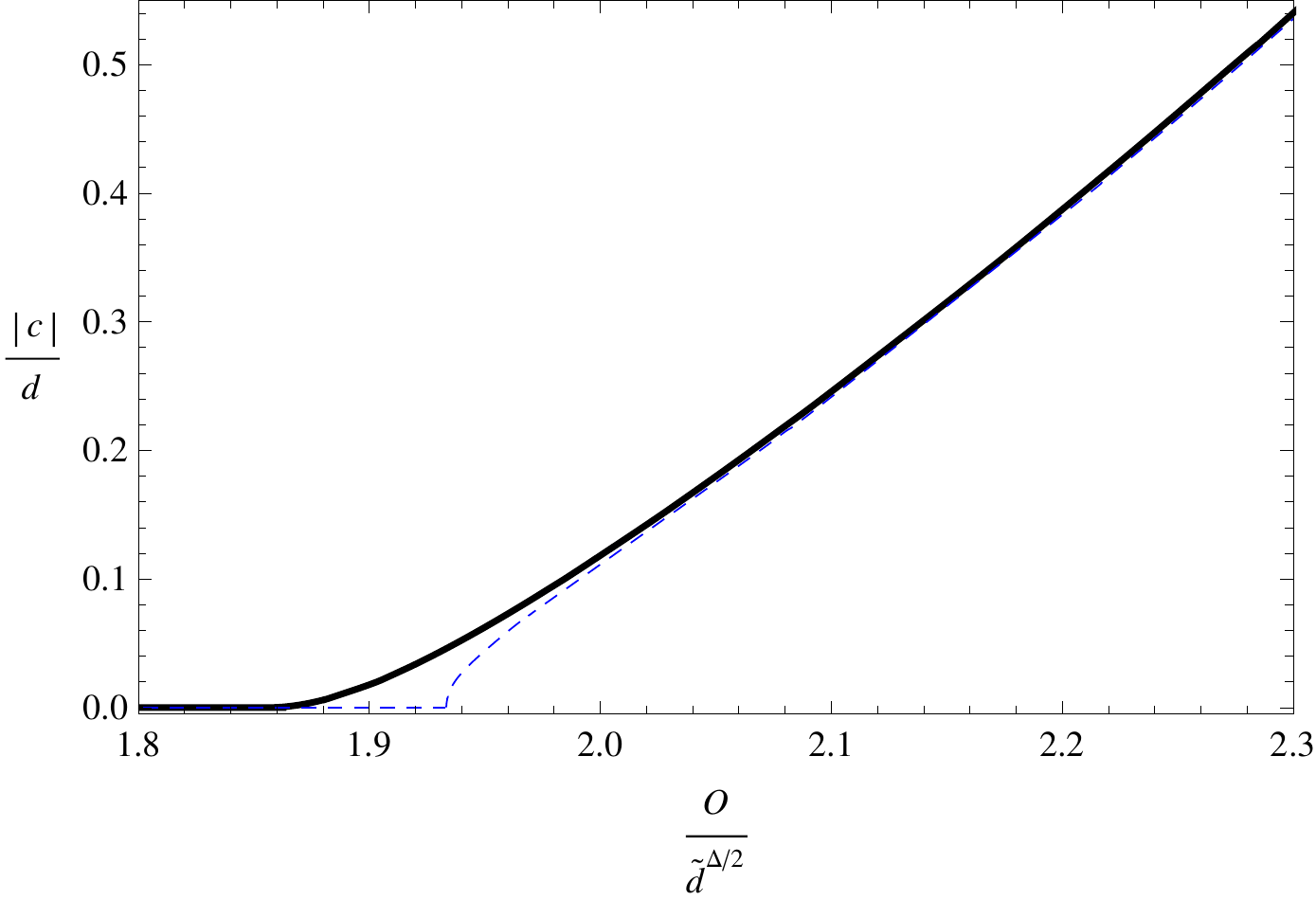}
  \caption{\label{HighT}The condensate in the deformed D3/D5
  system at zero (solid) and small (dashed)
  temperature $\pi T=10^{-5}\sqrt{\tilde{d}}$,
  large magnetic field $B=0.98B_c$, and the choice $\Delta=1$.
  The non mean-field scaling at zero temperature is destroyed even at
  this small temperature. The nonzero temperature condensate asymptotes
  to the zero temperature value far away from the transition.}
\end{figure}

We plot the condensate near the nonzero temperature transition
with small temperature $\pi T=10^{-5}\sqrt{\tilde{d}}$, magnetic
field $B=0.98B_c$, and the choice $\Delta=1$ in Fig.~\ref{HighT}.
At zero temperature and this magnetic field the condensate scales
with an exponent $\beta=2.18$, noticeably different from the
mean-field value obtained at the small temperature shown in the
figure.

\subsection{Fluctuations}
\label{flucs}

We now move on to consider dynamics. In particular  we will obtain
the structure of the retarded two-point function of the order
parameter at low frequency. This computation essentially mimics
the scaling and matching methods employed
in~\cite{Faulkner:2009wj}, so we will only quote the highlights.

We begin by considering a fluctuation of the worldvolume  field
$\theta$ around the symmetric embedding $\theta=0$. There is an
infinite tower of AdS$_4$ modes corresponding to the Kaluza-Klein
harmonics of $\theta$ reduced on the wrapped two-sphere; we only
consider the $s$-wave, as this is the lightest mode in the tower.
The chiral transition destabilizes it.

After a straightforward, if frustrating,  computation we obtain
the full Lagrangian for a time and spatially dependent embedding
$\theta=\theta(x^0,x^i,w)$. Denoting
\begin{equation}
\theta'=\partial_w \theta, \,\,\,\, \dot{\theta}=
\partial_{x^0}\theta, \,\,\,\,(\nabla \theta)^2 = (\partial_i \theta)(\partial_i\theta),
\end{equation}
where $i=1,2$, we find
\begin{align}
\nonumber
\tilde{\mathcal{L}}_5=-\mathcal{N}&\sqrt{1+\frac{1+
\frac{B^2}{w^4}}{1+\frac{B^2}{w^4}+\frac{(\nabla \theta)^2}{w^2}}
 \left( w^2f\theta'^2-\frac{\dot{\theta}^2}{w^2f} \right)} \\
&\times \sqrt{d^2+w^4\cos^4\theta\left( 1+\frac{B^2}{w^4}+\frac{(\nabla \theta)^2}{w^2} \right)}.
\end{align}
Now we must choose how to implement our deformation for the
system with a spatially-varying $\theta$. As before, we make the
choice that $O$ becomes a magnetic field for $\Delta=2$. This
amounts to taking
\begin{equation}
\frac{B^2}{w^4}\mapsto \frac{B^2}{w^4}+\frac{O^2}{w^{2\Delta}}.
\end{equation}

The two-point function of the condensate may be computed by
solving the bulk action to second order in the variation. That is,
by solving the linearized problem around $\theta=0$,
\begin{align}
\mathcal{D}\theta=0,
\end{align}
where $\mathcal{D}$ is a nasty second-order differential
operator. We also Fourier transform in the $x^{\mu}$ directions
and impose the incoming boundary condition at the horizon. The
resulting solution will have a near-boundary expansion
\begin{equation}
\theta=\frac{\theta_1(\omega,k)}{w}+\frac{\theta_3(\omega,k)}{w^3}+O(w^{-4}).
\end{equation}
Moreover, the two-point function of the condensate is  computed by
varying the regularized bulk action twice with respect to
$\theta_1$,
\begin{equation}
\label{greens}
G(\omega,k)=\langle \mathcal{O}(\omega,k)\mathcal{O}(-\omega,-k)\rangle
= K \frac{\theta_3(\omega,k)}{\theta_1(\omega,k)},
\end{equation}
for $K$ a positive constant. Solving for $G$ at  low energies is a
bit tricky as the correct infrared behavior of $\theta$ depends on
$\omega$ at leading order. We therefore solve $\theta$ in the
infrared region and match it to the outer region.

In order to solve for $\theta$ near the bottom of  the brane we
employ a scaling limit. Consider
\begin{equation}
\label{scale1}
w=\frac{\lambda}{\xi}, \,\,\, t=\lambda^{-1}\tau,
\end{equation}
in the $\lambda\rightarrow 0$ limit with $\xi,\tau$ finite.  At
zero temperature, the equation of motion of $\theta$ becomes that
of a $m^2=-2B^2/(\tilde{d}^2+B^2)$ scalar field in AdS$_2$,
\begin{equation}
\partial_{\xi}^2\theta+\left( \omega^2+\frac{2B^2}{\tilde{d}^2+B^2}\frac{1}{\xi^2} \right)\theta = 0.
\end{equation}
At nonzero temperature, however, we supplement the scaling limit Eq.~(\ref{scale1}) with
\begin{equation}
\label{scale2}
w_h = \frac{\lambda}{\xi_0}, \,\,\, \xi_0\text{ finite.}
\end{equation}
The equation of motion in the infrared is then
\begin{equation}
\label{scaleEoM}
\partial^2_{\xi}\theta+\frac{\partial_{\xi}h}{h}\partial_{\xi}\theta+
\left( \frac{\omega^2}{h^2}+\frac{2B^2}{\tilde{d}^2+B^2}\frac{1}{h\xi^2} \right)\theta=0,
\end{equation}
where
\begin{equation}
h=1-\frac{\xi_0^4}{\xi^4}.
\end{equation}
At zero temperature, the scaling limit Eq.~(\ref{scale1})  amounts
to the $\omega\ll \sqrt{\tilde{d}},\sqrt{B}$ limit. At nonzero
temperature Eq.~(\ref{scale2}) is the $\omega,T\ll
\sqrt{\tilde{d}},\sqrt{B}$ limit with $\omega\sim T$. Notably we
can solve for $\theta$ in this region at both zero and nonzero
temperature.  Even more importantly, these equations of motion are (i.) independent of both $O$ and its dimension, $\Delta$ and (ii.) those of a scalar in either AdS$_2$ or an AdS$_2$ space with a black hole.

Unfortunately the scaling limits Eq.~(\ref{scale1}),(\ref{scale2})
do not give rise to a systematic matching program. As noted
in~\cite{Faulkner:2009wj}, a proper matching divides the $w$ axis
into two regions
\begin{align}
&\textbf{Inner:   }w=\frac{\omega}{\xi}, \,\,\, \text{for}\,\,\, \xi\in (\epsilon,\infty), \\
&\textbf{Outer:   }\frac{\omega}{\epsilon}<w,
\end{align}
in the limits
\begin{equation}
\omega\rightarrow 0, \,\,\, \xi=\text{ finite}, \,\,\, \epsilon\rightarrow 0, \,\,\, \frac{\omega}{\epsilon}\rightarrow 0.
\end{equation}
Small $\omega$ perturbations can be treated systematically in
each region, employing $\xi$ as the coordinate in the inner one
and $r$ for the outer. The result has the form
\begin{align}
&\text{Inner:}\,\,\, \theta_I(\xi)=\theta_I^{(0)}(\xi)+\omega \theta_I^{(1)}(\xi)+\hdots \\
&\text{Outer:}\,\,\, \theta_O(r)=\theta_O^{(0)}(w)+\omega \theta_O^{(1)}(w)+\hdots.
\end{align}
The domain of these solutions overlaps in the region defined  by
$\xi\rightarrow 0$ with $w=\omega/\xi\rightarrow 0$; the full
solution is obtained by matching $\theta_I$ and $\theta_O$ there.

Now we solve for $\theta$ in the inner region. The leading  order
equation of motion for $\theta_I(\xi)$ is identical to the one we
found after the scaling limit, Eq.~(\ref{scaleEoM}). Near the
boundary of the AdS$_2$ region (that is, $\xi\rightarrow 0$), the
leading order term in $\theta_I$ can be expanded as
\begin{equation}
\theta_I^{(0)}(\omega,k,\xi)=\varphi_+(\xi)+\mathcal{G}_{\Delta_{\rm IR}}(\omega)\varphi_-(\xi),
\end{equation}
where $\varphi_{\pm}(\xi)$ are the non-normalizable/normalizable
solutions to Eq.~(\ref{scaleEoM}) and $\mathcal{G}_{\rm
IR}(\omega)$ is the retarded Green's function of the operator dual
to $\theta$ in the infrared theory. It takes on two vastly
different forms depending on whether we are at exactly zero or
nonzero temperature. For the first, it is~\cite{Faulkner:2009wj}
\begin{equation}
\mathcal{G}_{\Delta_{\rm IR}}(\omega)\propto (i\omega)^{2\Delta_{\rm IR}-1},
\end{equation}
while at nonzero temperature it is
\begin{equation}
\mathcal{G}_{\Delta_{\rm IR}}(\omega)\propto (i\omega)T^{2\Delta_{\rm IR}-1}.
\end{equation}
The precise form of $\mathcal{G}$ can be found in~\cite{Faulkner:2009wj}.

The bottom of the outer region corresponds to the near-boundary
region on the infrared AdS$_2$. The solution to $\theta_O$
therefore has the same functional form there, and so we can choose
a basis where the linearly independent solutions for $\theta_O$
match precisely to $\varphi_{\pm}$ in the infrared. That is,
\begin{equation}
\theta_O^{(0)}(w)=\eta^{(0)}_+(w)+\mathcal{G}_{\Delta_{\rm IR}}(\omega) \eta^{(0)}_-(w),
\end{equation}
where $\eta^{(0)}_{\pm}(w)$ is our (zeroth-order) basis in the
outer region. At higher order in $\omega$ the matching can be
systematically employed, effectively correcting the basis at each
order so that
\begin{equation}
\theta_O(w)=\eta_+(w)+\mathcal{G}_{\Delta_{\rm IR}}(\omega)\eta_-(w)
\end{equation}
is satisfied.

Near the AdS$_4$ boundary the $n^{th}$ order corrections to
$\eta_{\pm}$  will have an expansion
\begin{equation}
\eta^{(n)}_{\pm}(w)=\frac{a^{(n)}_{\pm}(\omega,k)}{w}(1+\hdots)+\frac{b^{(n)}_{\pm}(\omega,k)}{w^3}(1+\hdots).
\end{equation}
This together with Eq.~(\ref{greens}) leads to the desired
result, namely the form of the retarded two-point function
\begin{equation}
\label{twoPoint}
G(\omega,k)=K\frac{b_+^{(0)}+O(\omega)+\mathcal{G}_{\Delta_{\rm IR}}(\omega)(b_0^{(0)}+O(\omega))}{a_+^{(0)}+O(\omega)+\mathcal{G}_{\Delta_{\rm IR}}(a_-^{(0)}+O(\omega))}.
\end{equation}
Moreover, by expanding the $a$'s and $b$'s about $k=0$, we find
that $G$ assumes the small $\omega,k$ form
\begin{equation}
\label{twoPtLim}
G(\omega,k)\sim \frac{g_0+g_1(i\omega)^{2\Delta_{\rm IR}-1}+g_2k^2}{f_0+f_1(i\omega)^{2\Delta_{\rm IR}-1}+f_2k^2},
\end{equation}
as long as $\Delta_{\rm IR}<1$. For $\Delta_{\rm IR}>1$, the low-energy limit is instead
\begin{equation}
G(\omega,k)\sim\frac{\tilde{g}_0+\tilde{g}_1\omega+\tilde{g_2}k^2}{\tilde{f}_0+\tilde{f}_1\omega+\tilde{f}_2k^2}.
\end{equation}

On the reasonable assumption that the matching
coefficients, $g_i$ and $f_i$, are analytic in $O-O_c$, then the chiral transition corresponds to a root in $f_0$. That is, near the transition $f_0$ is proportional to $O_c-O$. We then find that for
$\Delta_{\rm IR}<1$ the mode that drives the transition obeys a
zero temperature dispersion relation
\begin{equation}
f_1(i \omega)^{2\Delta_{\rm IR}-1}+f_2k^2\propto O_c-O.
\end{equation}

From this relation we simultaneously obtain the dynamical
critical exponent $z$ at the transition,
\begin{equation}
z=\frac{2}{2\Delta_{\rm IR}-1},
\end{equation}
and the divergence of the correlation length,
\begin{equation}
\langle \phi(x)\phi(0)\rangle \sim e^{-|x|/\xi}, \,\,\, \xi\sim (O_c-O)^{-\nu}, \,\,\, \nu=\frac{1}{2}.
\end{equation}
Thus the dynamical exponent takes a non mean-field value while
$\nu$ takes the mean-field one. At nonzero temperature, however,
the dispersion relation becomes
\begin{equation}
f_0 (i\omega) T^{2\Delta_{\rm}-1}+f_1 k^2=O_c-O,
\end{equation}
so that the dynamical critical exponent takes on a mean-field value $z=2$. As with the condensate, the non
mean-field scaling is destroyed at any nonzero temperature.

\section{An Effective model}
\label{eff}

We have shown that in the D3/D5 system with a magnetic field,
density and a phenomenological operator $O$ there is a rich phase
structure. The chiral restoration transition is of the holographic BKT type at
large $B$ but second order at large $O$ with a region in between
with non mean-field behaviour. Our results provide a rich array of
numerical data that we will show can be completely matched by a
simple effective theory.

We\footnote{This result is based on unpublished work by KJ with
T. Faulkner.} have been able to guess the form of the effective
potential. 
For $\Delta_{\rm IR}>3/4$, the mean
field scaling of static exponents is reproduced by the potential
\begin{equation}
\label{effV} V_{eff}(\phi)=V_0 +
\alpha_2(O_c-O)\phi^2+\alpha_4\phi^4+O(\phi^6),
\end{equation}
where the couplings $\alpha_i$ are presumed to be
positive and $V_0$ is the free energy in the symmetric phase. This
generates the expectation value $\phi\sim \sqrt{O-O_c}$.
 This is just a standard Landau-Ginzburg model.

The crucial ingredient when we move away from mean-field scaling
is that the gravity dual reveals that the infra-red dynamics is
governed by a lower dimensional AdS$_{p+1}$ theory (for the probe brane systems, $p=1$). We have also learnt that the order parameter in this low energy regime acts as an operator of either dimension $\Delta_{IR}$
or $p-\Delta_{IR}$ (see Eq.~(\ref{dIR}) for the D3/D5 case).
It is not immediately clear which case holds true; however, we have tried each possibility and find success
with our effective model if the dimension of the condensing
operator is taken as $p- \Delta_{IR}$.

Now it is natural, on dimensional grounds, to include an
additional term in the potential for our order parameter $\phi$
coming from the $p$ dimensional theory
\begin{equation}
\label{effV} \Delta V_{eff}(\phi)=\alpha_{\rm
IR}\phi^{p/(p-\Delta_{\rm IR})},
\end{equation}
Again $\alpha_{\rm IR}$ is assumed to be positive.

If $ \frac{3}{4} p < \D_{IR} < p$,
then the quartic contribution to the effective potential dominates over the term from the infrared theory. In this case, minimizing the potential yields the standard mean-field critical exponent.
However, if $\frac{1}{2}p <
\D_{IR} < \frac{3}{4}p$, then we find the non-mean-field exponent
\begin{eqnarray}
  \phi_0 \sim (O-O_c)^{\beta}  \equiv (O-O_c)^{\frac{p-\Delta_{IR}}{2\Delta_{IR}-p}} \ . \label{exponent}
\end{eqnarray}
%
%
When the bound $\Delta_{\rm IR}<\Delta_c = 3/4p$ 
is satisfied, the condensate scales with a 
non mean-field exponent.

For the example of the magnetic field competing with our operator
$O$ in the D3/D5 system. We can find the expected critical
exponent
\begin{eqnarray}
  && \beta = \frac{1-\Delta_{IR}}{2\Delta_{IR}-1} = \half\left(\sqrt{\frac{\tilde{d}^2+B^2}{\tilde{d}^2-7B^2}} - 1   \right) \ , \nn \\
  &&  \left( \sqrt{\frac{3}{29}}\tilde{d} < B <  \sqrt{\frac{1}{7}}\tilde{d} \right) \label{analy}
\end{eqnarray}
where we set $p=1$ and the $B$ range comes from $1/2< \Delta_{IR}
< 3/4 $. As a result $1/2 < \beta < \infty$. For $B <
\sqrt{\frac{3}{29}}\tilde{d} $, $\beta = 1/2$. We have plotted the result Eq. (\ref{analy}) over our numerical results
in Fig.\ref{nmfExp} and they match the numerical results
perfectly.

In many ways Eq.~(\ref{effV}) is the primary result of this work.
The non-trivial emergent theory leaves a fingerprint on the
effective potential which can lead to rich phase diagrams even in
the strict $N\rightarrow\infty$ limit. For example, if
$\Delta_{\rm IR}<5/6$ then a non mean-field tri-critical point is
realized by driving either $\alpha_4$ (for $\Delta_{\rm IR}>3/4$)
or $\alpha_{\rm IR}$ (for $\Delta_{\rm IR}<3/4$) negative while
keeping the other positive. In the first case, the terminating
line of second-order transitions has mean-field exponents while in
the second it does not.

We can further test our effective potential by measuring the free
energy near the transitions we identified  in Sec.~\ref{nmfSec}.
if we write the effective potential as
\begin{equation}
\label{Veff1} V_{eff}(\phi)=V_0+\alpha_2(O_c-O)\phi^2+\alpha_{\rm
IR}\phi^{2+\frac{1}{\beta}},
\end{equation}
In the broken phase this gives
\begin{equation}
\phi\sim (O-O_c)^{\beta}, \,\,\,\, \Delta F\sim
(O-O_c)^{1+2\beta},
\end{equation}
giving a prediction for how the free energy should scale across
the transition.

\begin{figure}[]
\includegraphics[width=8cm]{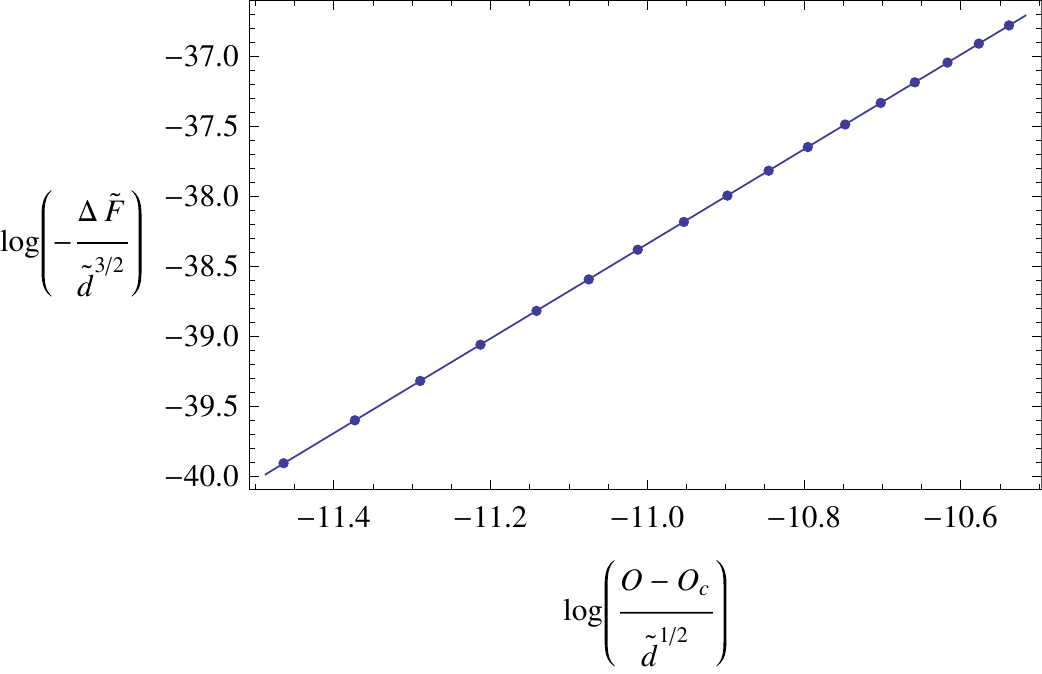}
\caption{\label{freeE}A log-log plot of the difference of free
energy,  $\Delta F$, in the broken phase at zero temperature as a
function of $O$. The dots indicate numerical data at magnetic
field $B=0.95 B_c$ and the choice $\Delta=1$, and the line a
numerical fit. The free energy scales as $\Delta F\sim
-(O-O_c)^{3.38}$ in the broken phase. }
\end{figure}

Recall that in order to measure the free energy, we compute
(minus) the regularized bulk action. We do this numerically,
employing the methods of~\cite{Jensen:2010vd}. We plot some
representative results at relatively large magnetic field,
$B=0.95B_c$ and the choice $\Delta=1$ in Fig.~\ref{freeE}.
Numerically, we find that the condensate scales with exponent
$\beta=1.20$ and the free energy in the broken phase as $\Delta
F/\mathcal{N}\sim -(O-O_c)^{3.38}$. which indeed reproduces the
scaling of the free energy for $\beta=1.20$ to within $1\%$.
We repeated this analysis for many different values of
$\Delta_{\rm IR}$  and found the same basic result. The scaling of
the free energy is reproduced by the effective potential
Eq.~(\ref{Veff1}) for the corresponding exponent $\beta$. This
shows the strength of the effective potential analysis which does
not need to know even the dimension of the operator $O$.

The effective potential Eq.~(\ref{effV}) can be generalized for
cases where the emergent theory has different scaling symmetries
than in the deformed D3/D5 system. The structure is essentially
the same: a Landau-Ginzburg potential analytic in $\phi^2$ and a
non-analytic piece stemming from the infrared theory. In the next
section we will show such an application to the D3/D7 system.

\section{BKT and Non Mean Field Transitions in the D3/D7 System}
\label{d3d7}

As another example of our phenomenological analysis and effective
theory methods let us finally return briefly to the D3/D7 system 
in Eq.(\ref{phenoS7}). That system has a magnetic field, density and a
phenomenological operator $O$ present.

Looking at linearized fluctuations around the symmetric $L=0$
embedding,  there are three different infrared limits depending on
the value of $\Delta$. At large $\Delta>3$, the field $L/\rho$
fluctuates unstably in the infrared. For $\Delta<3$, $L/\rho$
fluctuates as a stable massless scalar in AdS$_2$. However, for
the particular case $\Delta=3$, $L/\rho$ fluctuates as a
$m^2=-3O^2/(\tilde{d}^2+O^2)$ scalar in AdS$_2$. Adjusting the
ratio $O/\tilde{d}$ then tunes the dimension of the scalar
operator dual to $L$ in the infrared theory to be
\begin{equation}
\Delta_{\rm
IR}=\frac{1+\sqrt{\frac{\tilde{d}^2-11O^2}{\tilde{d}^2+O^2}}}{2}.
\end{equation}

This last case is perhaps the most interesting.  Notably, it
corresponds to the case where the density and simulated
deformation have the same dimension. As with the D3/D5 system at
nonzero magnetic field and density, increasing $O$ at fixed
density will trigger a chiral holographic BKT transition at
$\tilde{d}/O_c=\sqrt{11}$ as the field $L/\rho$ violates the BF
bound in the effective AdS$_2$ region. On the other hand, at
smaller values of $O$ we can presumably drive a second-order
chiral transition with the magnetic field. There will be a regime
of non mean field transitions for some intermediate values of $O$.
The choice $\Delta=3$ is therefore analogous to our
phenomenological D3/D5 system with the roles of the magnetic field
and $O$ reversed.

We can apply our effective field theory Eq. (\ref{effLG}) to this
case too. It predicts the critical exponents
\begin{eqnarray}
  && \beta =  \half\left(\sqrt{\frac{d^2+O^2}{d^2-11O^2}} - 1   \right) \ , \nn \\
  &&  \left( \sqrt{\frac{3}{43}}d < O <  \sqrt{\frac{1}{11}}d \right) 
\end{eqnarray}

\section{Discussion}
\label{discuss}

We now summarize our results. For our
phenomenological D3/D5 setup with magnetic field, density, and a
third control parameter we find the non-trivial phase diagram in
Fig.~\ref{phases}. The new ingredient is that tuning a third
control parameter can lead to a line of second-order transitions.
Moreover, the critical exponents of these transitions do not
appear to depend on the details of the deformation. Rather they
are functions of the dimension, $\Delta_{IR}$, of the operator dual to the
embedding function in an emergent infrared theory. This dimension
is tuned by the equal-dimension (in the UV theory) control
parameters density and magnetic field.

We have measured or computed four of the critical exponents  along
this line. In Sec.~\ref{nmfSec}, we numerically measured the
condensate in the broken phase and found agreement with an
analytic function of the infrared dimension as in
Eq.~(\ref{betaExp}). In terms of the magnetic field and density,
the exponent $\beta$ is given by Eq.(\ref{analy}). At smaller
$B/\tilde{d}$ (below $\sqrt{3/29}$), $\beta$ is simply $1/2$. We also computed the free
energy in the broken phase and found that both its scaling and
$\beta$ follow from an effective potential Eq.~(\ref{effLG}). From
this we can also compute the critical exponent $\gamma$, which is
related to the scaling of the susceptibility,
\begin{equation}
\frac{\partial \phi}{\partial m}\sim (O-O_c)^{-\gamma},
\end{equation}
since the hyper mass $m$ is conjugate to the condensate $\phi$.
The effective potential with a mass becomes
\begin{equation}
V_{m,eff}(\phi)=V_{eff}(\phi)+m\phi.
\end{equation}
The exponent $\gamma$ is computed from the effective  potential to
be the mean-field value $\gamma=1$.

At nonzero temperature, however, all non mean-field  scaling is
lost and the effective potential becomes an ordinary quartic
polynomial. The exponents $\beta$ and $\gamma$ are $1/2$ and $1$
respectively. The temperature destroys the non mean-field scaling,
just as in the holographic BKT transitions.

We also computed the low-energy behaviour of the two-point
function of the condensate in Sec.~\ref{flucs}. Fluctuations of
the condensate correspond to time and spatially-dependent
fluctuations of the bulk field $L$. At small $w\ll
\sqrt{B},\sqrt{\tilde{d}}$, the equation of motion for $\theta\sim
L/\rho$ resembles that of a scalar in AdS$_2$. The equation is
solvable there and we match it to the physics outside the infrared
region. This enables us to show that at $T=0$ the dynamical 
critical exponent is non mean-field,
\begin{equation}
\nn
z=\frac{2}{2\Delta_{\rm IR}-1}
\end{equation}
dependent only upon $\Delta_{\rm IR}$. Moreover it returns to
the mean field value $z=2$ at finite temperature.

In the end, our main results are dependent upon the details of
the emergent infrared theory. As a result we expect them to hold
in a much wider class of problems, including conformal theories
with three control parameters as discussed. It would be nice to
test this picture in a purely field theoretic context. In
particular, it would be extremely interesting to realize both the
emergent theory as well as the sort of phase diagram we identify
in a conformal theory without a holographic dual.

\acknowledgements  K.J. is pleased to thank Andreas Karch and Carlos
Hoyos for their insights and conversation. He also thanks Thomas Faulkner for related collaboration to this work.
K.J. was supported, in part, by the U.S. Department of Energy
under Grant Numbers DE-FG02-96ER4095. K-Y.K. was supported  by STFC.
N.E. is, in part, supported by STFC and an IPPP Associateship.

\bibliography{refs}

\end{document}